\documentclass[aps,pra,showpacs,twoside,twocolumn,longbibliography,10pt]{revtex4-1}
\usepackage[colorlinks=true, citecolor=red, urlcolor=blue ]{hyperref}
\usepackage{epsfig,newlfont,amssymb,amsfonts,amsmath,bm,subfigure,palatino,mathtools,amsthm,braket,times,soul,enumitem,color}
\usepackage[normalem]{ulem}
\newcommand{\stkout}[1]{\ifmmode\text{\sout{\ensuremath{#1}}}\else\sout{#1}\fi}
\usepackage[english]{babel}
\usepackage[utf8]{inputenc}
\usepackage{array}
\usepackage{xcolor}
\usepackage{graphics}

\def\Tr{\text{Tr}}

\usepackage{verbatim}
\usepackage{bbm}
\usepackage{wrapfig}

\usepackage{hyphenat}

\newlength\figureheight 
\newlength\figurewidth

\begin{document}

\title{Krylov Complexity and Quantum Many-body Scars in \\ Non-Hermitian Kinetically Constrained Models}
\author{Arkaprava Sil, 
Rudraksh Kaushik, 
Souvik Kundu, 
Prasant Mallik, 
Sudipto Singha Roy
}

\affiliation{Department of Physics, Indian Institute of Technology (ISM) Dhanbad, Dhanbad 826004, India
}

\begin{abstract}
Thermalization in closed quantum systems and its breakdown are central to understanding the emergence of statistical behavior in isolated many-body systems. Kinetic constraints can hinder thermalization by restricting the allowed dynamics and giving rise to non-ergodic behavior. While various aspects of ergodicity breaking in kinetically constrained models have been extensively studied in Hermitian systems, their behavior in non-Hermitian kinetically constrained models remains less explored. In particular, the inequivalence of left and right eigenstate ensembles in non-Hermitian systems raises the question of whether they exhibit qualitatively different dynamical and eigenstate properties. In this work, we study non-Hermitian generalizations of the paradigmatic PXP and PPXPP models, focusing on their Krylov complexity, eigenstate structure, and dynamics. By exploiting the Arnoldi's algorithm, we find distinct behavior of Krylov complexity in the left and right eigenstate ensembles as non-Hermiticity is increased: in the right ensemble, it exhibits persistent oscillations with an increasing period, while in the left ensemble, it gradually saturates. The characteristic arch-shaped profile of the Arnoldi coefficients is also modified differently in the two ensembles, with a Hermitian-like arch re-emerging in the left ensemble at large non-Hermiticity. Additionally, in the non-Hermitian PXP model, the right ensemble retains the characteristic scar structure, while the left ensemble develops additional non-scar eigenstates with anomalously low entanglement and large overlap with the vacuum state.  These findings reveal that non-Hermiticity can affect Krylov dynamics and eigenstate structure differently in the left and right ensembles, demonstrating the need to treat them separately when characterizing non-Hermitian kinetically constrained systems.
\end{abstract}

\maketitle

\section{Introduction}

Understanding thermalization and quantum complexity in any generic nonintegrable isolated quantum many-body system is crucial from both a theoretical and an application-based perspective. The eigenstate thermalization hypothesis (ETH)~\cite{ETH_ansatz1, ETH_ansatz2, ETH_ansatz3, ETH_ansatz4, ETH_ansatz5, ETH_ansatz6, ETH_ansatz7} has been proven to be successful  in explaining thermalization phenomena in such systems. 
However, a breakdown of thermalization is also often observed in interacting many-body systems and itself constitutes a different study. Although integrability~\cite{integrability_1, integrability_2} and many-body localization~\cite{MBL_1, MBL_2, MBL_3, MBL_4, MBL_5} provide mechanisms for strong ergodicity breaking, there are also known mechanisms leading to weak ergodicity breaking, notably quantum many-body scars (QMBS)~\cite{scar_1, scar_2, scar_3, PPXPP} and Hilbert-space fragmentation~\cite{HSF_1, HSF_2, HSF_3, Sil_2026}. 
In quantum many-body scars, some highly excited states in the energy spectrum deviate from the ETH property. These states typically exhibit anomalously low entanglement entropy~(see also~\cite{PPXPP}), and can give rise to pronounced fidelity revivals, signaling the persistence of memory of the initial state.  In Hilbert-space fragmented systems, kinetic constraints divide the total Hilbert space into many dynamically disconnected sectors, commonly known as Krylov subspaces. Consequently, global thermalization is precluded, and thermalization, if there is any, takes place independently within individual Krylov sectors~\cite{Moudgalya2021}. \\

 Although these phenomena differ in their microscopic origins, they share a common feature: the dynamics is constrained from exploring the full available Hilbert space. This naturally motivates studying how the dynamics spreads within the accessible space and how such spreading differs between thermalizing, weakly ergodicity breaking,  and nonthermalizing systems. In a generic chaotic nonintegrable quantum many-body system, the dynamics progressively spreads quantum information over the accessible Hilbert space, leading to the loss of local memory of the initial state and, under suitable conditions, thermalization. 
In this regard, Krylov complexity~\cite{Krylov_1, Krylov_2, Krylov_3, Krylov_4, Krylov_5,Krylov_6,krylov_7} has emerged as a useful framework to quantify such spreading and the complexity associated with the dynamics of many-body systems. While out-of-time-ordered correlators (OTOCs)~\cite{Lin_2018, Zanardi_2021, Suchetan_2022_1, Suchetan_2022_2, Xu_2024, Jisha_2026}, spectral statistics~\cite{Haake_2010}, and other diagnostics~\cite{Serbyn_2017,Dowling_2024} have been widely employed to characterize the complexity and chaotic nature of a system, Krylov complexity provides a direct dynamical measure of the spreading of a state or operator in the Krylov basis, and the resulting Krylov-space distribution allows one to quantify both the extent and rate of this dynamical spreading, making it particularly useful for probing constrained and nonergodic dynamics. This naturally raises the question of how Krylov complexity behaves in systems exhibiting weak ergodicity breaking, where dynamical spreading can remain constrained despite the absence of conventional integrability. In particular, quantum many-body scars provide a natural setting in which such anomalous dynamics can arise: when an initial state has a significant overlap with the scarred subspace, its evolution can retain memory of the initial state and exhibit dynamics distinct from that of a typical state in a chaotic system. Recent works in this direction have demonstrated that Krylov complexity can capture signatures of weak ergodicity breaking and reveal characteristic features of the dynamics associated with quantum many-body scars~\cite{Budhaditya_2022, Qingmin_2025,Caputa_2025}.

In this regard, most of the studies of thermalization, its breakdown, and Krylov complexity have focused on isolated Hermitian quantum many-body systems. Realistic quantum systems, however, can deviate from this idealized setting due to gain and loss~\cite{Eleuch_2017}, asymmetric hopping~\cite{Yao_2018, Shen_2024}, external driving~\cite{Ke_2023, Tista_2024, Cai_2024, Dash_2026}, or coupling to an environment~\cite{Link_2017, Ahuja_2026}, leading naturally to non-Hermitian descriptions. While Hermiticity was traditionally imposed to ensure real energy eigenvalues, it is now well established that non-Hermitian systems can also possess entirely real spectra under appropriate conditions. A prominent example is the $\mathcal{PT}$-symmetric phase, where the spectrum remains real as long as the symmetry is unbroken ~\cite{Bender_Hook_2023}. Moreover, non-Hermitian systems exhibit a range of phenomena with no direct Hermitian counterpart, including exceptional points~\cite{Heiss_2004, Bergholtz_2021, Loran_2022, Ronika_2023, Quinn_2023, Dash_2026, Agarwal_2026_1}, non-Hermitian skin effect~\cite{Yao_2018, Lee_2019, Wang_2023, Aditi_2023, Padhi_2024},  which have motivated extensive studies of their spectral and dynamical properties. In particular, the enhanced response of non-Hermitian systems near exceptional points has attracted considerable interest for quantum sensing and precision measurements~\cite{Wiersig_2014, Wiersig_2016, Agarwal_2026_2}. These developments highlight the potential of non-Hermitian many-body systems not only as a platform for studying unconventional quantum dynamics, but also for applications in quantum technologies.

In this direction, Ref.~\cite{nonherm_ETH} provides a framework for investigating thermalization in non-Hermitian systems, while Refs.~\cite{Krylov_open_1, Krylov_open_2, Krylov_open_3, Krylov_open_4, Krylov_open_5, Krylov_open_6} have explored Krylov complexity in  open quantum systems. Despite these developments, the interplay between Krylov complexity and weak ergodicity breaking in non-Hermitian systems~\cite{Chen_2023, Shen_2024}  is still not fully understood. In particular, a key feature distinguishing non-Hermitian systems from their Hermitian counterparts is the inequivalence of the left and right eigenstate ensembles, which can exhibit markedly different physical properties~\cite{Soumik_2025, Lakkaraju_2026}.  This raises the natural question of whether such left-right asymmetry is also reflected in Krylov-space dynamics and in the structure of atypical or scarred eigenstates. Understanding these aspects can therefore provide a more complete picture of weak ergodicity breaking and dynamical complexity in non-Hermitian many-body systems.

In this article, we investigate weak ergodicity breaking in non-Hermitian many-body systems for the non-Hermitian PXP and PPXPP models, which naturally arise in Rydberg atom arrays due to the blockade constraint. Our study addresses two complementary aspects of the resulting non-Hermitian dynamics: the spreading of quantum information in Krylov space and the structure of quantum many-body scarred and atypical eigenstates. While these two aspects are physically related through anomalous dynamical behavior, we initially treat them as distinct diagnostics and subsequently establish a connection between them. For the Krylov analysis, we independently construct the left and right Krylov bases using the Arnoldi~\cite{Arnoldi_1951} algorithm and examine how the corresponding Krylov complexity changes with the non-Hermitian deformation. Separately, we investigate the atypical eigenstates in the left and right eigenstate ensembles, focusing on the evolution of scarred states and their entanglement properties away from the Hermitian limit.  

Our results reveal a pronounced asymmetry between the two eigenstate ensembles. The Krylov complexity exhibits qualitatively different behavior in the left and right sectors as the non-Hermitian parameter is varied, demonstrating that the spreading of quantum information in Krylov space is sensitive to the left-right structure of the non-Hermitian problem. A corresponding asymmetry is also found in the eigenstate structure. While the known scarred states display distinct behavior in the two ensembles, we additionally identify atypical low-entanglement states in the left ensemble that do not possess the characteristic signatures of quantum many-body scars. Thus, low entanglement alone does not provide a sufficient criterion for identifying scars in the non-Hermitian setting.\\

Taken together, our results demonstrate that the left and right eigenstate ensembles can encode substantially different dynamical and structural information, and that neither their Krylov complexity nor their atypical-state structure can, in general, be inferred directly from the Hermitian limit. Our work therefore highlights the necessity of treating the two ensembles independently when characterizing non-Hermitian many-body dynamics and weak ergodicity breaking.\\

The article is organized in the following way: In Sec.~\ref{sec: Models}, we describe the models used in our study. We also set the basic nomenclature and convention concerning non-Hermitian systems. In Sec.~\ref{sec:formalism}, we review the Krylov complexity and Arnoldi's algorithm that we have used. In Sec.~\ref{sec: results}, we present our obtained results in the non-Hermitian PXP model that includes Krylov complexity analysis, behavior of quantum many-body scars and other atypical states in left and right ensembles of the model. Finally, in Sec.~\ref{sec:discussion}, we discuss the importance of our findings and possible future directions.

\section{Models and formalism}
\label{sec: Models}

We start the discussion by introducing  the physical models of which we are going to study the non-Hermitian version, namely, the PXP~\cite{scar_1} and the PPXPP model~\cite{PPXPP}. The PXP model is an effective model that describes a one-dimensional (1D) Rydberg atom array subjected to the Rydberg blockade condition. Mathematically, the Hamiltonian  of the system can be expressed as follows
\begin{eqnarray}
    H_{\text{PXP}} = \sum_{j=1}^L P_{j-1}\sigma^x_jP_{j+1}, 
\end{eqnarray}
where $P_k=|0\rangle\langle0|_k$ is the projector onto the $|0\rangle$ state at the $k$-th site, and $\sigma^x_k$ is the Pauli matrix in the $x$ direction acting on the $k$-th site.  
 Here,  and in all the models considered in this work, we apply periodic boundary conditions (PBC). This model is known to host certain atypical eigenstates known as quantum many-body scars that do not thermalize~\cite{scar_1}. These eigenstates have large overlap with the N{\'e}el state $|Z_2\rangle = |1010...\rangle$ and show significantly lower entanglement entropy than other states in the bulk of the spectrum of the model. Since the number of such states is very low compared to the all eigenstates, and it scales linearly with the size of the system, $L$, the PXP model is said to break ergodicity weakly. Due to  the large overlap with scar eigenstates, when the dynamics is probed using the $|Z_2\rangle$ state, the  time evolved  fidelity shows revival-- counter to what is expected in generic quantum many-body chaotic systems. 

The non-Hermiticity in this model can be introduced by generalizing the symmetric coupling to be asymmetric, i.e. $|1\rangle\langle0| + |0\rangle \langle 1| \mapsto (1+\gamma)|1\rangle\langle0| + (1-\gamma)|0\rangle\langle1|$\footnote{In the QuSpin package~\cite{QuSpin}, the computational basis states are conventionally defined as $|0\rangle=(0,1)^{\mathsf T}$ and $|1\rangle=(1,0)^{\mathsf T}$. In all our numerical results, we have followed this convention}. Using this, we get the form of the non-Hermitian PXP model ~\cite{Chen_2023}-
\begin{eqnarray}
    H_{\text{NH-PXP}} = \sum_{j=1}^L P_{j-1} (\sigma^x_j + i\gamma \sigma^y_j) P_{j+1}.
\end{eqnarray}
 Here, $\gamma$ is called the non-Hermiticity strength and all other quantities are the same as the Hermitian case ($\gamma=0$). For $|\gamma|<1$, this system has real eigenvalues. We point out here that the non-Hermitian PXP model can be related to the Hermitian PXP model through a non-unitary global transformation. To see this, we write $\sigma^x + i\gamma\sigma^y$ as $(1+\gamma)\sigma^+ + (1-\gamma)\sigma^-$. Now, defining $S= e^{\theta \sigma^z/2}$, we get 
\begin{align}
    S(\sigma^x+i\gamma\sigma^y)S^{-1}= (1+\gamma)e^{\theta} \sigma^+ + (1-\gamma)e^{-\theta} \sigma^- .
\end{align}

\noindent In order to have both coefficients the same, we arrive at the following condition on $\theta$:   $e^{\theta} = \sqrt{\frac{1-\gamma}{1+\gamma}}$. Using this, we get the following.
\begin{align}
    S(\sigma^x+i\gamma\sigma^y)S^{-1} = \sqrt{1-\gamma^2} \ \sigma^x.
\end{align}

\noindent Defining the global operator $\mathcal{S} = \prod_i e^{\theta \sigma^z_i/2}$, which  commutes with all the $P_k$'s one can finally  yield
\begin{align}
    \mathcal{S} H_{\text{NH-PXP}} \mathcal{S}^{-1} = \sqrt{1-\gamma^2} H_{\text{PXP}}.
\end{align}

\noindent Now if $H_{\text{PXP}}$ has eigenvalues $\{E_i\}$ with eigenvectors $|E_i\rangle$, then $H_{\text{NH-PXP}}$ has eigenvalues $\{\sqrt{1-\gamma^2} E_i\}$ with right eigenvectors $\mathcal{S}^{-1}|E_i\rangle$ and left eigenvectors $\langle E_i| \mathcal{S}$, up to proper normalization. Since the spectrum of $H_{\text{PXP}}$ is real, for $|\gamma|<1$, $H_{\text{NH-PXP}}$ also shows real eigenvalues, and for $|\gamma|>1$, it shows completely imaginary eigenvalues given by the same relation.\\

The PPXPP model~\cite{PPXPP} can be thought of similarly to the PXP model but with an extended Rydberg blockade radius. In the PXP model, each excitation is surrounded by at least one ground state configuration on both sides. In the case of the PPXPP model, each excitation must be surrounded by atleast two ground state configurations on both sides. Therefore, $|1\rangle$ can appear in a state only if it appears as $|...00100...\rangle$. This model is also known to host quantum many-body scar states, resulting in weak ergodicity breaking. Similarly to the non-Hermitian PXP model, we can also define the non-Hermitian PPXPP model, given by 
\begin{eqnarray}
    H_{\text{NH-PPXPP}} = \sum_{j=1}^L P_{j-2}P_{j-1} (\sigma^x_j + i\gamma \sigma^y_j) P_{j+1} P_{j+2}.
\end{eqnarray}

\noindent We discuss the findings related to the  PPXPP model in Appendix~\ref{PPXPP}.\\
\noindent Both of these models have translation (under periodic boundary condition) and spatial inversion symmetry. If $\mathcal{T}$ is the translation operator, then $\mathcal{T}|\psi\rangle = e^{ik}|\psi\rangle$, where $k$ is the momentum, the generator of the translation operator. Since the lattice size is $L$, we must have $\mathcal{T}^L = I$. This gives $e^{ikL} = e^{i 2\pi n}$. So, $k=\frac{2\pi n}{L}$ ($n$ is any integer number). Fixing $n$ gives different momentum sectors. For example, $n=0$ gives the translationally invariant zero momentum sector.  Throughout this work, we have used the appropriate momentum sector wherever required and explicitly mentioned this in the corresponding sections. The action of the spatial inversion operator $\mathcal{I}$ is to take an inversion of the lattice about the mid-point. It maps the $i$th lattice site to the $(L-i+1)$th site. Consequently, $\mathcal{I}^2 = I$,  and its eigenvalues are $\pm1$. The eigenvalue $+1$ corresponds to the inversion-symmetric sector, while $-1$ corresponds to the inversion-antisymmetric sector. However, in the present work, we have not resolved the inversion symmetry.

\noindent Before proceeding further, we briefly discuss and set our conventions for non-Hermitian systems. Let us assume that $H_{\mathrm{NH}}$ is a non-Hermitian Hamiltonian ($H_{\mathrm{NH}} \neq H_{\mathrm{NH}}^{\dagger}$). For a Hermitian Hamiltonian, the right and left eigenstates are identical. Therefore, once the right eigenstate is known, the corresponding left eigenstate can be obtained simply by taking its Hermitian conjugate. Also, the Hermiticity of Hamiltonian ensures that the energy eigenvalues are real. However, for a non-Hermitian Hamiltonian, the energy eigenvalues in general can be complex, and the corresponding right and left eigenstates are generally different from each other. 
They satisfy the eigenvalue equations
\begin{eqnarray}
    H_{\mathrm{NH}}|R_i\rangle = E_i |R_i\rangle, \ \ \ \ H_\mathrm{NH}^{\dagger}|L_i\rangle = E_i^* |L_i\rangle. 
\end{eqnarray}
\noindent Here, $\{|R_i\rangle\}$ and $\{|L_i\rangle\}$ denote the sets of right and left eigenstates, respectively, which we also refer to as the right and left eigenstate ensembles. In the Hermitian case, since $|R_i\rangle = (\langle L_i|)^\dagger$, we simply denote the eigenstates by $|E_i\rangle$.

One of the primary goals of our work is to investigate the differences between the ensembles generated by the left and right eigenvectors of these kinetically constrained models. While the biorthogonal basis provides the natural framework for the spectral description of a non-Hermitian Hamiltonian, it does not imply that the properties of the left and right eigenstate ensembles are identical. In particular, starting from the same initial state, $\rho_0$, the two evolutions
\[
\rho_R(t)=e^{-iH_{\mathrm{NH}}t}\rho_0 e^{iH_{\mathrm{NH}}^\dagger t},
\qquad
\rho_L(t)=e^{-iH_{\mathrm{NH}}^\dagger t}\rho_0 e^{iH_{\mathrm{NH}}t},
\]
can be substantially different, resulting in distinct return probabilities, entanglement growth, and complexity dynamics. This motivates us to study the left and right eigenstate ensembles separately using a range of complementary diagnostics.

\section{Krylov complexity in non-Hermitian  model}
\label{sec:formalism}
 The first diagnostic employed to characterize the left and right eigenstate ensembles is the Krylov complexity, which quantifies the spreading of an initially localized state in the Krylov basis under many-body Hamiltonian dynamics. In the present work, we extend this framework to non-Hermitian systems, where the right and left Krylov spaces are constructed separately using the Arnoldi algorithm.
Before going into the detailed discussion, we first briefly review the formalism of Krylov complexity for Hermitian systems and then generalize it to our case. 

Let us  assume that our initial state is $|\phi(0)\rangle$, which is subjected to time evolution under the Hermitian Hamiltonian $H$, \begin{eqnarray}
    |\phi(t)\rangle = e^{-iHt}|\phi(0)\rangle.
\end{eqnarray} Now using the Lanczos algorithm~\cite{Lanczos1950}, we build a set of orthonormal basis vectors $\{|K_n\rangle\}_{n=0}^{K-1}$ from the set of states $\{|\phi(0)\rangle, H|\phi(0)\rangle, H^2|\phi(0)\rangle, ...\}$. Here, $\{|K_n\rangle\}$ is known as Krylov basis vectors and $K$ is denoted as the Krylov dimension. In the Lanczos algorithm, we typically orthogonalize each newly generated basis vector with the basis vectors generated in the previous two steps,  which automatically makes it orthogonal with all the previous basis vectors. If we write the Hamiltonian $H$ in this Krylov basis, it takes a tridiagonal form  whose eigenvalues provide increasingly accurate approximations to the Hamiltonian spectrum as the number of Lanczos iterations increases. 
Expanding the above state $|\phi(t)\rangle$  in the Krylov basis in the following way, we get 
\begin{eqnarray}
    |\phi(t)\rangle &=& \left( \sum_{n=0}^{K-1} |K_n\rangle\langle K_n|\right) |\phi(t)\rangle, \\ \nonumber
    &=& \sum_{n=1}^{K-1} \phi_n(t) |K_n\rangle, 
\end{eqnarray}
where $\phi_n(t) = \langle K_n|\phi(t)\rangle$ is the expansion coefficient in the Krylov basis. Now the Krylov spread complexity is defined as ~\cite{Krylov_2}-
\begin{eqnarray}
    C_K(t) = \sum_{n=0}^{K-1} n \ |\phi_n(t)|^2.
\end{eqnarray}

\noindent However, ambiguities arise when our Hamiltonian  becomes non-Hermitian, $H_{\mathrm{NH}}$. Traditional Lanczos algorithm fails to produce the tridiagonal Hamiltonian that reproduces the eigenvalues of the original non-Hermitian $H_{\mathrm{NH}}$. In such cases, one  uses either the Bi-Lanczos algorithm  or the Arnoldi's algorithm. The Bi-Lanczos method~\cite{Saad1980} creates two sets of biorthogonal Krylov vectors $\{|K_m^R\rangle\}$ and $\{|K_m^L\rangle\}$ for the right and left ensembles, respectively, using $\mathcal{K} = \text{Span}\{ |\phi\rangle, H_{\mathrm{NH}}|\phi\rangle, H_{\mathrm{NH}}^2|\phi\rangle,...\}$ and $\mathcal{K}^* = \text{Span}\{|\phi\rangle, H_{\mathrm{NH}}^{\dagger}|\phi\rangle, (H_{\mathrm{NH}}^{\dagger})^2|\phi\rangle, ...\}$, respectively. These two sets of basis satisfy biorthonormality condition: 
\begin{eqnarray}
    \langle K_m^L|K_n^R\rangle = \delta_{mn}, \ \ \ \langle K_m^{R(L)}|K_n^{R(L)}\rangle \neq \delta_{mn}.
\end{eqnarray}
\noindent If we arrange the right Krylov basis vectors $\{|K_m^R\rangle\}$ and the left Krylov basis vectors $\{|K_m^L\rangle\}$ as the columns of the matrices $P$ and $Q$, respectively,  then $Q^{\dagger}H_{\mathrm{NH}}P$ transforms the Hamiltonian $H_{\mathrm{NH}}$ into a tridiagonal matrix that reproduces the correct eigenvalues. \\

\noindent Arnoldi's method~\cite{Arnoldi_1951} is another way to deal with the non-Hermitian Hamiltonians. It reduces to the traditional Lanczos algorithm for Hermitian Hamiltonians. Unlike the traditional Lanczos algorithm, where making the newly generated basis vector orthogonal with only the last two basis vectors is sufficient, in Arnoldi's case, the condition is stringent and one needs to orthogonalize it with all the previously generated basis vectors. To find the basis vectors, in this case, one  needs to choose a probe vector $|U_0\rangle$. Below we  summarize the steps to find the set of basis states: 
\begin{enumerate}
    \item Initialize  $|V_0\rangle = |U_0 \rangle/|| \ |U_0\rangle ||$
    \item $|U_k\rangle = H|V_{k-1}\rangle$
    \item For $i$ in range $0$ to $K-1$:\\
           $h_{i, k-1} = \langle V_i|U_k\rangle$ \\
          $ \ \ \ \ |U_{k}\rangle = |U_k\rangle - \sum_{i=1}^{k-1} h_{i, k-1} |V_i\rangle $ \\
          $ \ \ \ \ h_{k, k-1} = || \ |U_k\rangle \ ||$\\
          If $h_{k, k-1} =0 :$ Break\\
          Else: $|V_k\rangle = |U_k\rangle/h_{k, k-1}$
    
\end{enumerate}
\noindent The resulting set of vectors $\{|V_k\rangle\}$ forms the desired orthonormal basis. 

In our analysis, we employ the Arnoldi method to construct the right and left Krylov bases of the non-Hermitian Hamiltonian and investigate the corresponding Krylov spread complexity. The Arnoldi algorithm is applied to $H_{\mathrm{NH}}$ and $H^{\dagger}_{\mathrm{NH}}$ to generate the right and the left Krylov basis vectors, $\{|V_m^R\rangle\}$ and $\{|V_m^L\rangle\}$, respectively. Let $R$ and $S$ be the matrices whose columns are the right and the left Krylov basis vectors, $\{|V_m^R\rangle\}$ and $\{|V_m^L\rangle\}$, respectively.  Then the transformed matrices $\tilde{H}_{\mathrm{NH}}=(R^{\dagger}H_{\mathrm{NH}}R)$ and $\tilde{\tilde{H}}_{\mathrm{NH}}=(S^{\dagger}H_{\mathrm{NH}}S)$ both produce the correct eigenvalues of the Hamiltonian $H_{\mathrm{NH}}$. However, in these cases, the transformed Hamiltonian is not a tridiagonal Hamiltonian, rather, $\tilde{H}_{\mathrm{NH}}$, which takes the upper-Hessenberg form which looks like the following : 

\begin{eqnarray}\tilde{H}_{\mathrm{NH}}=
\begin{pmatrix}
h_{11} & h_{12} & h_{13} & \cdots & h_{1n} \\
h_{21} & h_{22} & h_{23} & \cdots & h_{2n} \\
0      & h_{32} & h_{33} & \cdots & h_{3n} \\
\vdots & \vdots & \vdots & \ddots & \vdots \\
0      & \cdots & 0      & h_{n,n-1} & h_{nn}
\end{pmatrix},
\end{eqnarray}
where $h_{i,j} = \langle V_i|U_{j+1}\rangle = \langle V_i|H|V_{j}\rangle$, as used in the Arnoldi algorithm to find the basis vectors. In the conventional Lanczos algorithm, $a_n$ and $b_n$ denote the diagonal and sub-diagonal Lanczos coefficients, respectively. It is evident from the Arnoldi algorithm that $h_{i,i}$ and $h_{i,i-1}$ are analogous to $a_n$ and $b_n$, respectively. Since these coefficients encode important information about the dynamical properties of the system, we likewise investigate the behavior of $h_{i,i}$ and $h_{i,i-1}$ in the present work. For convenience, we also denote them as $a_n$ and $b_n$, respectively. Here, the left and right basis vectors $\{|V_m^L\rangle\}$ and $\{|V_m^R\rangle\}$ do not follow the biorthonormality condition; rather, they separately satisfy the orthonormality condition:
\begin{eqnarray}
    \langle V_m^L|V_n^R\rangle \neq \delta_{mn}, \ \ \ 
    \langle V_m^{R(L)}|V_n^{R(L)}\rangle = \delta_{mn}.
    \label{eq:arnoldi_orthonormal}
\end{eqnarray}
Using these normalized right basis vectors, we expand the time-evolving state in the following form:
\begin{equation}
|\phi(t)\rangle=\sum_n \phi_n^{(R)}(t)|V_n^{(R)}\rangle,
\end{equation}
where $\phi_n^{(R)}(t)$ are the Krylov-space amplitudes for the right ensemble. We then calculate the Krylov complexity \begin{eqnarray}
    C_K^{(R)}=\sum_{n=0}^{K-1}n |\phi_n^{(R)}(t)|^2, 
    \end{eqnarray}
    from these amplitudes, following the same procedure discussed above. We call this the Krylov complexity in the right eigenstates ensemble. A similar procedure is also followed with the left eigenstates, yielding Krylov complexity in the left ensemble (denoted by $C_K^{(L)}$). 

In this context, we remind the reader again that   one of the objectives of this work is to compare the Krylov complexities obtained from the left- and right eigenstate ensembles. For this purpose, we construct the corresponding left and right Krylov bases separately, with each basis being orthonormal within itself. This provides a convenient framework for carrying out the comparison on an equal footing and motivates our use of the Arnoldi method in this work.

Another distinction between the Arnoldi and Bi-Lanczos approaches concerns the expansion of the time-evolved state. In the Bi-Lanczos method, the time-evolved state can be expanded in terms of the biorthogonal Krylov bases as
\begin{eqnarray}
|\phi(t)\rangle &=& \left(\sum_n  |K_n^R\rangle \langle K_n^L| \right) \ |\phi(t)\rangle, \\ \nonumber
&=& \sum_n \phi_n^L(t) \ |K_n^R\rangle .
\end{eqnarray}
In this case, since the left and right Krylov bases are biorthogonal rather than orthonormal individually, the expansion coefficients generally do not satisfy $\sum_n |\phi_n^L(t)|^2 = 1$. If required, these coefficients can therefore be normalized before evaluating quantities based on their probability distribution. In contrast, as Eq.~(\ref{eq:arnoldi_orthonormal}) suggests, the Krylov basis obtained from the Arnoldi procedure is orthonormal, and hence the expansion coefficients directly satisfy the usual normalization condition when the time-evolving state is normalized. This provides a convenient way to define and compare the Krylov-space probability distributions for the left and right ensembles considered here.

\section{Results}
\label{sec: results}
With the background information in hand, we are now ready to elaborate on our main findings. Below, we first discuss the Krylov complexity analysis and later compare the quantum many-body scar states appearing in the left and right ensembles, highlighting their significant differences from their Hermitian counterparts.
\subsection{Krylov complexity in non-Hermitian PXP Model}

\begin{figure*}
    \centering
    \includegraphics[width= 0.9\textwidth]{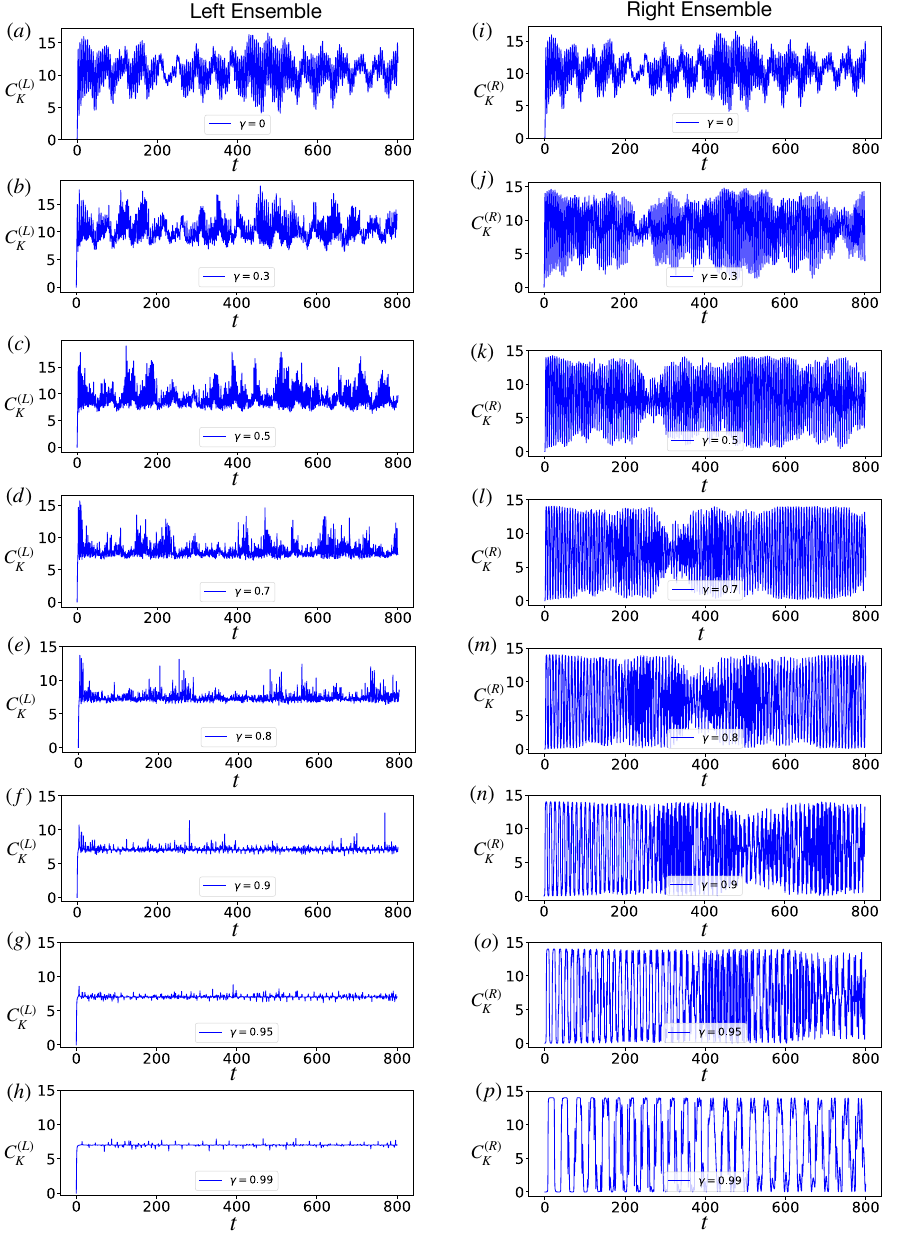}
   \caption{{\bf Krylov complexity for the left and right ensembles in the non-Hermitian PXP model with increasing $\boldsymbol{\gamma}$.} In (a)--(h), we show the behavior obtained for the left ensemble ($C_K^{(L)}$) for $\gamma=0.0,\,0.3,\,0.5,\,0.7,\,0.8,\,0.9,\,0.95,\,0.99$, respectively, with $|Z_2\rangle$ taken as the initial state. The corresponding results for the right ensemble ($C_K^{(R)}$) are shown in (i)--(p). All plots are for $L=14$. For the left ensemble, increasing $\gamma$ leads to a progressive suppression of the oscillations and a tendency towards saturation of $C^{(L)}_K$, whereas for the right ensemble the oscillations in $C^{(R)}_K$ become increasingly prominent. This distinct behavior highlights the different Krylov-complexity dynamics of the left and right eigenstate ensembles in the non-Hermitian regime. One can observe that, for the left ensemble, $C^{(L)}_K$ saturates at a value that is almost half of the maximum value obtained for the right ensemble.}
    \label{fig:NHPXP_KC_both}
\end{figure*}

\begin{figure*}[htbp]
    \centering
    \includegraphics[width=\textwidth]{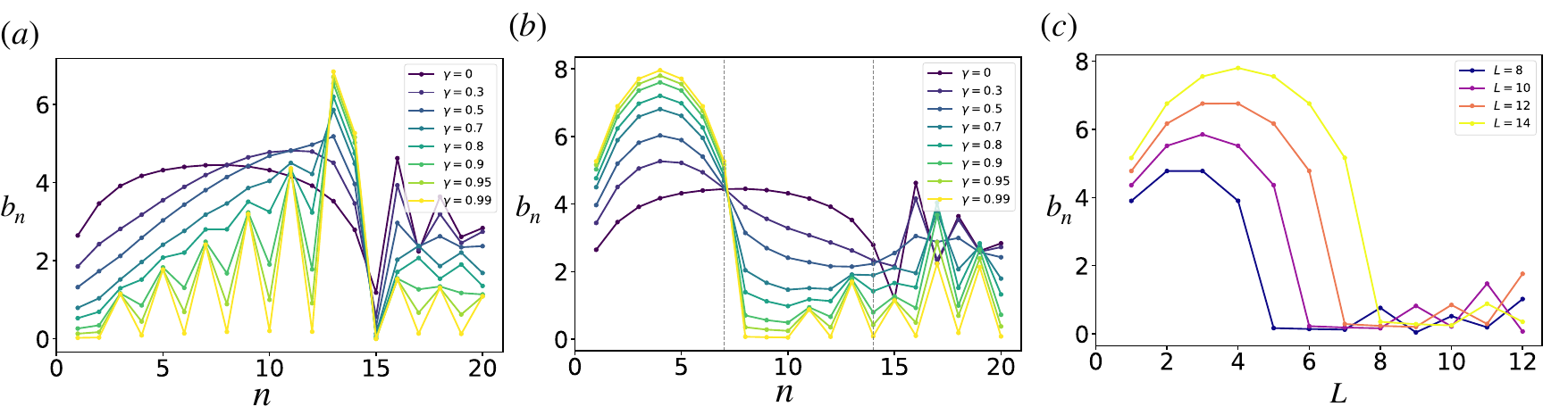}
\caption{\textbf{Behavior of Arnoldi coefficients with Krylov steps.} 
     (a) Arnoldi coefficients $b_n$ as a function of the Krylov 
    index $n$ for the right eigenvector ensemble, shown for various 
    non-Hermitian parameter values $\gamma$. The $\gamma = 0$ curve 
    corresponds to the Hermitian PXP model and exhibits the characteristic 
    arch-shaped profile associated with quantum many-body scar states. 
    As $\gamma$ increases, this arch-shaped structure is progressively 
    modified, revealing a behavior distinctly different from the Hermitian 
    limit. 
    (b) $b_n$ versus $n$ for the left eigenvector ensemble at the same 
    values of $\gamma$. Unlike the right ensemble, a Hermitian-like 
    arch-shaped imprint re-emerges even at large $\gamma$, but with a 
    notable $\gamma$-independent fixed point at $n = L/2$, marked by the 
    first vertical dashed line.  All panels are shown for $L = 14$.
    (c) $b_n$ as a function of system size $L$ for different chain lengths 
    ($L = 8, 10, 12, 14$) at $\gamma=0.95$, illustrating the system-size dependence of the 
    $\gamma$-independent  point in the left ensemble.}
    \label{fig:bn_plots}
\end{figure*}

Krylov complexity in Hermitian PXP model has already been studied in ~\cite{Caputa_2025}. Here, we briefly review them for comparison purposes with the non-Hermitian counterparts. As mentioned before, in our work, we use the Arnoldi's algorithm which for the Hermitian case naturally reduces to the Lanczos algorithm. The system size considered here is $L=14$ and we use $|Z_2\rangle = |1010...\rangle$ as the seed vector of the algorithm. In Fig.~\ref{fig:NHPXP_KC_both}(a) (or (i)), we have shown the dynamics of the Krylov complexity  measure for the conventional PXP model. From the plot, we can clearly see an imperfect oscillatory behavior in  $C_K(t)$. In ~\ref{fig:bn_plots}(a) (or (b)), the $\gamma=0$ plot corresponds to the  behavior of a few $b_n$s  with  $n$. We see that for some initial values, the Lanczos coefficients $b_n$ exhibit an arch-shaped profile that extends approximately up to $n=14$, which coincides exactly with the number of nonzero scar eigenstates present in this model for $L=14$. In Refs.~\cite{Budhaditya_2022,Caputa_2025}, it has been argued that the appearance of such an arch can be understood from a toy model with an approximately equally spaced tower of quantum many-body scar states, for which the Krylov coefficients can be obtained analytically and exhibit a similar arch-shaped structure. Physically, the scar tower provides a restricted subspace in which the dynamics is highly coherent, and the approximately equal energy spacing leads to an effective SU(2)-like structure in the corresponding Krylov dynamics. As a result, the Lanczos coefficients $b_n$, which describe the hopping between neighboring Krylov states, first increase, reach a maximum, and then decrease, giving rise to the characteristic arch. This provides a connection between the spectral structure of the scar states and the spreading of complexity in Krylov space. In particular, the extent of the arch is related to the size of the scar-dominated Krylov subspace, with a larger number of participating scar states leading to a broader structured region of $b_n$. These observations establish a direct connection between Krylov complexity and quantum many-body scars, which we investigate further in the non-Hermitian paradigm.  

The Arnoldi algorithm here yields a tridiagonal matrix representation of the PXP Hamiltonian, with all the diagonal elements $a_n$ vanishing. This can be understood from the bipartite structure of the PXP model. Starting from a product state $|V_0\rangle$, the states generated by an even number of applications of $H_{\mathrm{PXP}}$ belong to one set, while those generated by an odd number of applications belong to another set. Since $H_{\mathrm{PXP}}$ connects only states belonging to different sets, it has no matrix elements within the same set. Consequently, the diagonal matrix elements of the Krylov Hamiltonian vanish, giving $a_n=0$.

Interestingly, in the non-Hermitian case, notable differences arise in many of such behaviors reported above. 
In general, for  the open quantum system paradigm, where the dynamics is governed by Liouvillian (which is non-unitary), Arnoldi's algorithm yields an upper Hessenberg matrix in which the diagonal elements are purely imaginary, while the first superdiagonal elements $h_{i,i+1}$ and first subdiagonal elements $h_{i,i-1}$ are real. Furthermore, moving away from the first superdiagonal, the higher-order off-diagonal elements exhibit an alternating pattern of purely imaginary and real values. 
In contrast, for the non-Hermitian extension of our model, we find that all diagonal elements $h_{i,i}$ of the Hessenberg matrix vanish. Additionally, alternating diagonals in the upper-triangular part also vanish. These features can be understood from the chiral symmetry of the Hamiltonian. Starting from a state belonging to one chiral sector, successive applications of $H_{\mathrm{NH}}$ generate Krylov vectors that alternate between the two chiral sectors, while $H_{\mathrm{NH}}$ connects only states belonging to opposite sectors. Consequently, matrix elements of $H_{\mathrm{NH}}$ between Krylov vectors belonging to the same chiral sector vanish, which leads to the observed pattern of zero diagonal and alternating upper-triangular elements in the Hessenberg matrix. Therefore, in our case, the upper Hessenberg matrix obtained from the Arnoldi procedure is also completely real. 

We plot the dynamics of the Krylov complexity measure $C^{(L)}_K$ for the left ensemble in Figs.~\ref{fig:NHPXP_KC_both}(b)--(h), while the corresponding behavior for the right ensemble ($C^{(R)}_K$) is shown in Figs.~\ref{fig:NHPXP_KC_both}(j)--(p) for increasing values of $\gamma$. These results can be compared with the corresponding Hermitian case shown in Figs.~\ref{fig:NHPXP_KC_both}(a) and (i), which represent the limit $\gamma=0$. We observe that even for a small value of $\gamma$, a significant change in the behavior of Krylov complexity measure can be clearly observed. For the left ensemble, the oscillations in $C^{(L)}_K$ become weaker as $\gamma$ increases. At large $\gamma$, $C^{(L)}_K$ approaches a saturation value with only small fluctuations. However, for the right ensemble, the Krylov complexity continues to exhibit persistent oscillations with time even at large $\gamma$. In fact, stronger non-Hermiticity leads to more pronounced revivals. 
A possible explanation for this behavior comes from the structure of the eigenstates of the model in the non-Hermitian limit. As $\gamma\rightarrow1$, $|Z_2\rangle$ becomes an exact right eigenstate of the model and has a large overlap with the scar states. Thus, the right-ensemble dynamics becomes increasingly controlled by a small number of scar states. The interference between these scar states can then produce a clear oscillatory behavior in the Krylov complexity. If the scar states have an approximate energy separation $\Delta E_{\mathrm{scar}}$, then the oscillatory part 
yields a characteristic period
\begin{equation}
T_K\simeq\frac{2\pi}{\Delta E_{\mathrm{scar}}}.
\end{equation}
Therefore, if the relevant scar-state energy gap decreases with increasing $\gamma$, the oscillation period of $C_K^{(R)}$ will increase, or equivalently its frequency will decrease. This is consistent with the numerical observation. The left ensemble behaves differently because the corresponding left-state overlaps do not select the same small set of scar states in the strongly non-Hermitian regime, resulting in very low overlap with $|Z_2\rangle$. Rather, as we show in the following sections, a large number of eigenvectors in the left ensemble show considerable overlap with the state $|\Omega\rangle=|000\dots0\rangle$.  Consequently, the coherent interference responsible for the regular oscillations is not preserved in the same way, leading to a suppression of the oscillatory part and an approach toward a saturation value. These overlaps are explicitly shown and discussed in the following section, where we examine the quantum scars in detail.

\begin{figure*}[htbp]
    \centering
    \includegraphics[width=\textwidth]{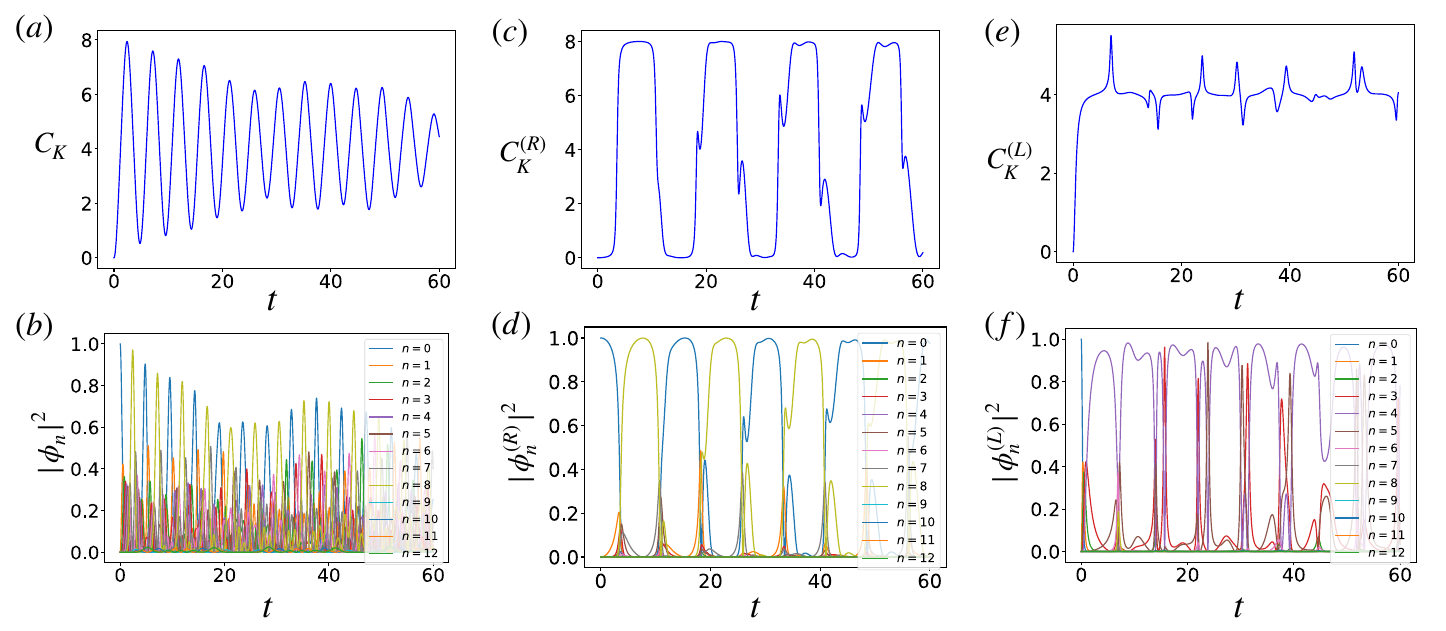}
   \caption{
\textbf{Krylov complexity and probability dynamics in Hermitian and non-Hermitian PXP systems.}
(a) We replot the Krylov complexity $C_K(t)$ as a function of time for the Hermitian PXP model, to compare with the corresponding Krylov-space probabilities $|\phi_n(t)|^2$. Here, we consider $L=8$, for which the Krylov-space dimension is $12$. 
(b) Shows the Krylov-space probabilities $|\phi_n(t)|^2$ for the Hermitian case for all $12$ Krylov states, highlighting the probability transfer between the Krylov states and the dominant oscillations of $|\phi_0(t)|^2$ and $|\phi_8(t)|^2$. 
(c) Depicts the Krylov complexity $C^{(R)}_K(t)$ for the non-Hermitian right ensemble, showing the modified oscillatory behavior of the complexity, and (d) supports this behavior through the corresponding dynamics of the right-ensemble Krylov probabilities $|\phi_n^{(R)}(t)|^2$ for all $12$ Krylov states. In particular, like the Hermitian case, here also the dynamics is predominantly governed by the oscillations of $|\phi_0^{(R)}(t)|^2$ and $|\phi_8^{(R)}(t)|^2$, with the latter providing the dominant contribution to $C^{(R)}_K(t)$ and yielding the approximate relation $C_K(t)\simeq 8|\phi_8^{(R)}(t)|^2$. 
(e) The Krylov complexity $C^{(L)}_K(t)$ for the non-Hermitian left ensemble illustrates the suppression of the pronounced oscillations observed for the right ensemble and the tendency towards saturation, which is further supported by (f), where we plot the corresponding left-ensemble Krylov probabilities $|\phi_n^{(L)}(t)|^2$ for all $12$ Krylov states. The latter shows a broader redistribution of probability across the Krylov basis and a comparatively large contribution from $|\phi_4^{(L)}(t)|^2$, which results in the suppression of the oscillatory complexity and its saturation around $C^{(L)}_K(t)\simeq 4$.
}
    \label{fig:phi_KC_all3}
\end{figure*}

To further understand the distinctive behavior of the complexity measure of the two ensembles, we investigate the behavior of the probability amplitudes $|\phi_n|^2$ contributing to  the Krylov complexity. We have plotted Krylov complexity and $\{|\phi_n|^2\}$ for the Hermitian system (Figs.~\ref{fig:phi_KC_all3}(a)-(b)), for the non-Hermitian right ensemble, $\{|\phi^{(R)}_n|^2\}$ (Figs.~\ref{fig:phi_KC_all3}(c)-(d)), and for the left ensemble, $\{|\phi^{(L)}_n|^2\}$ (Figs~\ref{fig:phi_KC_all3} (e)-(f)),  for a smaller system size, $L=8$, for which the Krylov space dimension is $12$. Here, we have plotted all the $12$ amplitudes. In Fig.~\ref{fig:phi_KC_all3}(b), we see that $|\phi_0|^2$ and $|\phi_8|^2$ show the highest amplitude in the oscillation with the same envelope, which has also been already reported in Ref.~\cite{Caputa_2025}. In the non-Hermitian right ensemble, this behavior persists: $|\phi^{(R)}_0|^2$ and $|\phi^{(R)}_8|^2$ are the only amplitudes that exhibit enhanced oscillations, while the oscillations in all other $|\phi^{(R)}_n|^2$ are strongly suppressed. Therefore, in this case, the Krylov complexity is approximately dominated by the contributions from $n=0$ and $n=8$,
\begin{equation}
C^{(R)}_K(t)\approx \sum_{n=0,8}n|\phi^{(R)}_n|^2
=0\times|\phi^{(R)}_0|^2+8\times|\phi^{(R)}_8|^2
=8|\phi^{(R)}_8|^2.
\end{equation}
Consequently, the temporal behavior of $C^{(R)}_K(t)$ in Fig.~\ref{fig:phi_KC_all3}(c) closely follows that of $|\phi^{(R)}_8|^2$ in Fig.~\ref{fig:phi_KC_all3}(d), with its amplitude enhanced by a factor of $8$. For the left ensemble, shown in Fig.~\ref{fig:phi_KC_all3}(f), we see that $|\phi^{(L)}_4|^2$ has a relatively large probability. Whenever $|\phi^{(L)}_4|^2$ decreases, the probability is transferred to some of the other Krylov states. As a result, the contributions from different Krylov states compensate each other, suppressing the oscillations in the Krylov complexity and leading to a saturation around $C^{(L)}_K\approx 4$.  

In this regard, we would like to mention another interesting observation. The Krylov complexity of the right ensemble for non-Hermiticity strength $+\gamma$ is exactly the same as that of the left ensemble for $-\gamma$. Therefore, if we replace $\gamma$ by $-\gamma$ in Fig.~\ref{fig:NHPXP_KC_both}, panels (a)--(h), which show the right ensemble, become the corresponding results for the left ensemble, while panels (i)--(p) become the corresponding results for the right ensemble. This behaviour can be understood directly from the form of the Hamiltonian. The non-Hermitian PXP Hamiltonian is given by
\begin{eqnarray}
    H(+\gamma)_{\text{NH-PXP}} = \sum_{j=1}^L P_{j-1} (\sigma^x_j + i\gamma \sigma^y_j) P_{j+1}.
\end{eqnarray}
Under the transformation $\gamma\rightarrow -\gamma$, we obtain the following.
\begin{eqnarray}
    H(-\gamma)_{\text{NH-PXP}} &=& \sum_{j=1}^L P_{j-1} (\sigma^x_j - i\gamma \sigma^y_j) P_{j+1}, \nonumber\\
    &=& H(+\gamma)_{\text{NH-PXP}}^{\dagger}.
\end{eqnarray}
Since the right eigenstates of $H_{\text{NH-PXP}}$ are related to the left eigenstates of $H_{\text{NH-PXP}}^\dagger$, and vise versa, changing $\gamma$ to $-\gamma$ exchanges the roles of the right and left eigenstates. Therefore, the behaviour of the right ensemble at $+\gamma$ is identical to that of the left ensemble at $-\gamma$, and vice versa.

\begin{figure*}[htbp]
    \centering
    \includegraphics[width=0.7\textwidth]{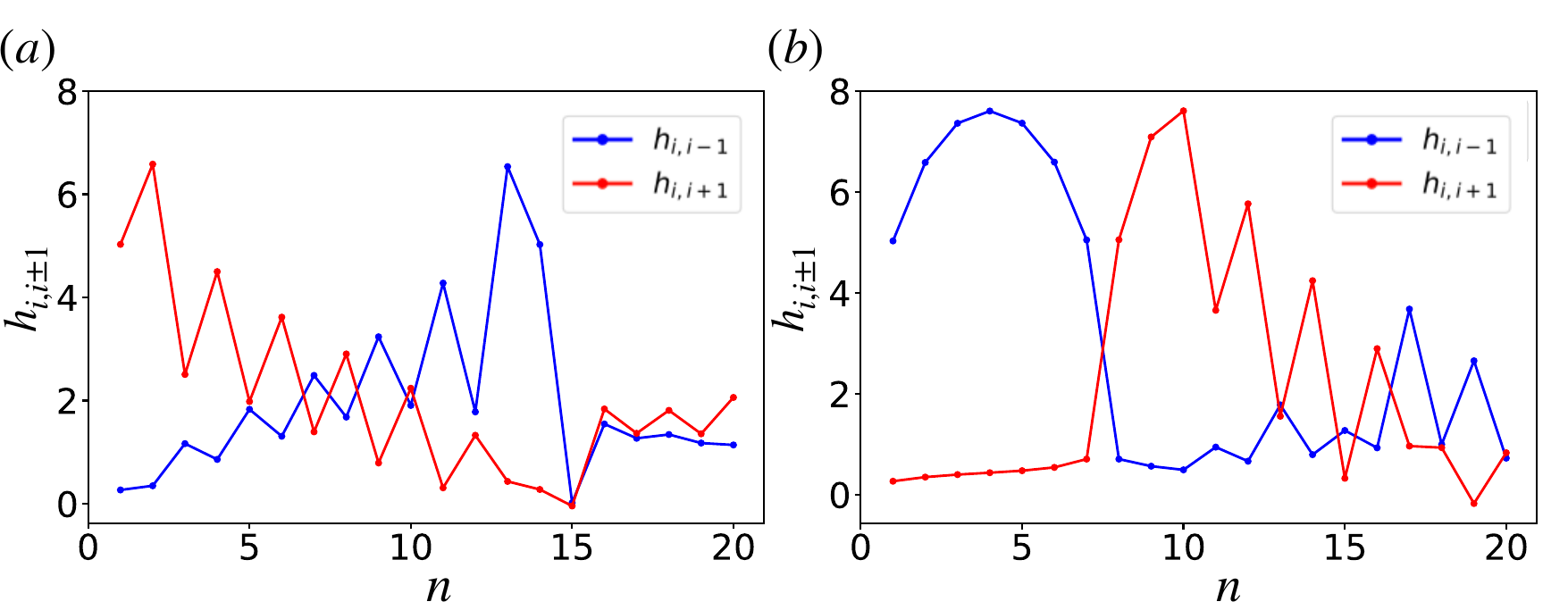}
   \caption{\textbf{Coefficients of the Hessenberg matrix in the right and left ensembles.}
(a) Here, we plot the  coefficients of the Hesseneberg matrix, $h_{i,i-1} (b_n)$ and $h_{i,i+1}$, as a function of the Krylov index $n$ for the right ensemble, while (b) shows the corresponding behavior for the left ensemble. In both cases, we observe that the behavior of $h_{i,i-1}$ is almost the opposite of  $h_{i,i+1}$.}
    \label{fig:extended_arch}
\end{figure*} 

We now revisit the $b_n$ versus $n$ plots shown in Fig.~\ref{fig:bn_plots} for the non-Hermitian case. As discussed above, the $\gamma=0$ curves in both panels (a) and (b) correspond to the Hermitian PXP model, and in this limit, the Lanczos coefficients show an arch-shaped profile, which is associated with the presence of quantum many-body scar states and reflects their contribution to the Krylov dynamics.  All nonzero values of $\gamma$ in Figure~\ref{fig:bn_plots}(a) present the corresponding results for the ensemble on the right eigenstates. As $\gamma$ increases, the arch-shaped structure observed in the Hermitian limit is gradually modified, and the behavior becomes distinctly different from that in the Hermitian case. In the next section, we investigate the fate of the non-Hermitian scar states in this ensemble, which may provide a characteristic diagnostic for understanding the observed behavior of the Krylov complexities and related quantities. Figure~\ref{fig:bn_plots}(b), on the other hand, shows that a Hermitian-like imprint reappears for the left ensemble even at a fixed large value of $\gamma$. However, unlike the Hermitian case, the arch-shaped structure shows a $\gamma$ independent point at  $n=L/2$. This behavior is observed for different spin-chain lengths $L$ (see Figure~\ref{fig:bn_plots}(c)), allowing us to examine how the position of this revival changes with the size of the system.

To further examine the fate of this arch in the non-Hermitian regime, we also consider the first superdiagonal elements $h_{i,i+1}$ of the upper Hessenberg matrix, in addition to the subdiagonal elements $h_{i,i-1}\equiv b_n$, for both the left and right ensembles. In the Hermitian limit, these two sets of coefficients are related by Hermiticity and hence contain equivalent nearest-neighbor information in Krylov space. In the non-Hermitian case, however, they are generally independent, allowing us to examine how the scar-related arch structure is modified along the two directions of Krylov-space connectivity. Figure~\ref{fig:extended_arch} shows that, for both ensembles, the behaviors of $h_{i,i-1}$ and $h_{i,i+1}$ remain complementary to each other. For instance, in the region where $h_{i,i+1}$ exhibits a sub-arch structure for the left ensemble, $h_{i,i-1}$ remains almost flat. 

\begin{figure*}[htbp]
    \centering
    \includegraphics[width=0.85\textwidth]{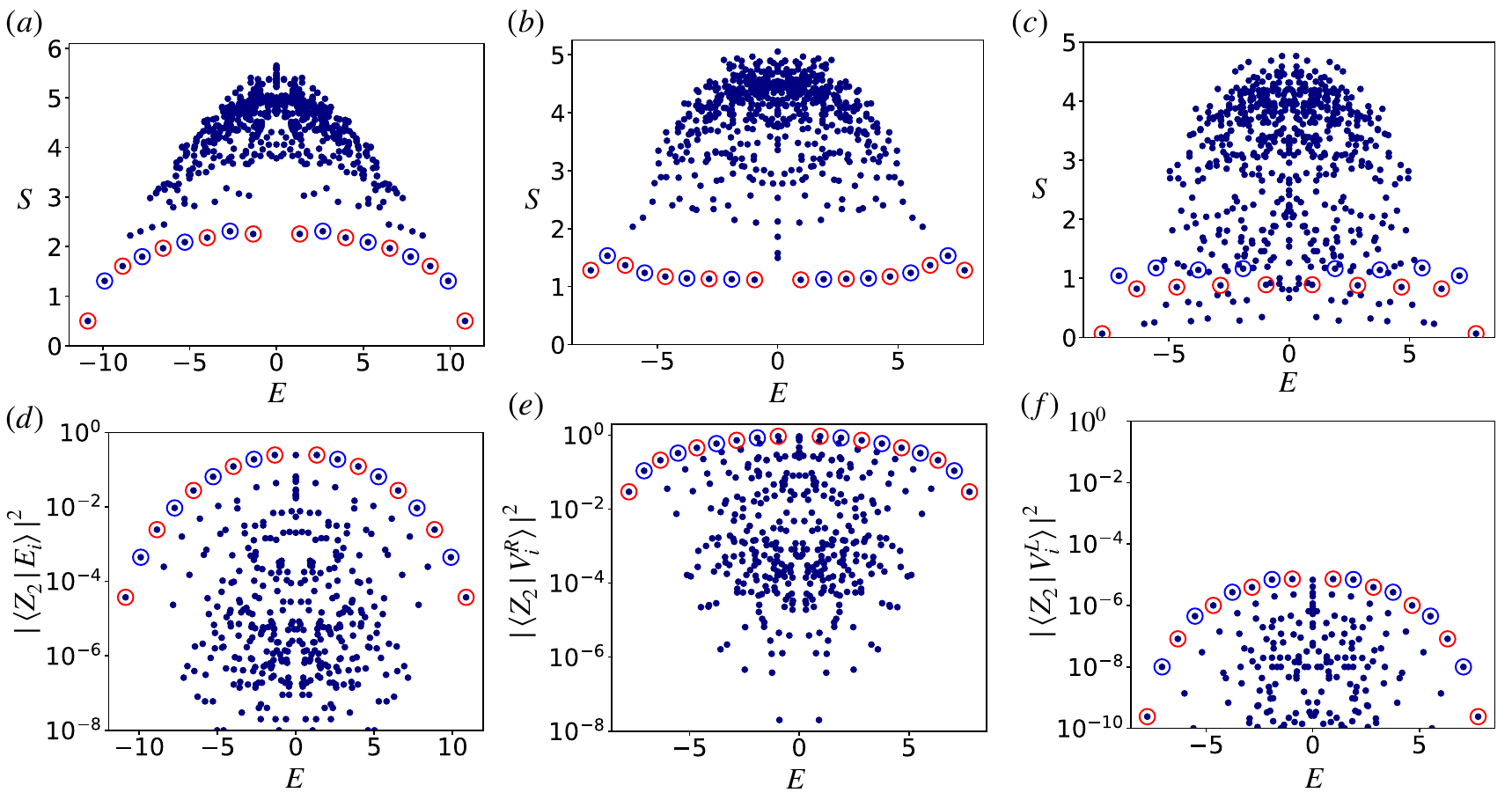}
\caption{{\bf Entanglement structure and scar-state overlaps in the Hermitian and non-Hermitian PXP models.} (a) Shows the behavior of entanglement entropy (EE) with energy $E$, for the Hermitian PXP model $(\gamma=0)$. (b) and (c) show the corresponding EE versus $E$ plots for the right and left ensembles, respectively, of the non-Hermitian PXP model at $\gamma=0.7$. (d) Overlap of the PXP eigenstates with the $|Z_2\rangle$ state. (e) and (f) show the corresponding overlaps for the right and left ensembles, respectively, of the non-Hermitian PXP model. All plots are for $L=18$ and include the two momentum sectors $k=0$ and $\pi$. The scar states are highlighted by circles, with red and blue circles denoting the $k=0$ and $k=\pi$ momentum sectors, respectively. In the left ensemble, the scar states in the two momentum sectors form distinct band-like structures in the EE versus energy plot in (c). They also exhibit stronger entanglement suppression than the corresponding right eigenstates, as seen in (b).}
    \label{fig:NHPXP_scar}
\end{figure*}

In the next section, we turn to the analysis of quantum many-body scar states and investigate their behavior in both ensembles. We then explore how the properties of these scar states can help us relate and understand some of the behavior observed in the Krylov complexity.

\subsection{Quantum Many-body Scars in non-Hermitian PXP model}

The conventional PXP model is a paradigmatic example of a nonintegrable many-body system hosting quantum many-body scar states. These atypical eigenstates are embedded within an otherwise thermal spectrum and are characterized by anomalous properties, such as low entanglement entropy and enhanced overlaps with product initial states. Consequently, they can strongly influence the dynamics and give rise to long-lived coherent oscillations, despite the nonintegrability of the underlying model. The revival in the Krylov complexity discussed above can be directly attributed to the contribution of such scar states, as we elaborate further in this section.

Recent studies have shown that quantum many-body scars persist in the non-Hermitian extension of the PXP model, with scarred eigenstates identified in both the right and biorthogonal ensembles~\cite{Chen_2023}. Here, in addition to these ensembles, we investigate the left ensemble and examine the presence and properties of its scarred eigenstates. In particular, we characterize the scar states using their entanglement entropy and overlaps with appropriate reference states, and compare their properties across the different ensembles.

\subsubsection{Entanglement Entropy as a marker of QMBS}
To investigate quantum scars, one of the common tools used is the entanglement entropy. The entanglement entropy of a pure state $|\psi\rangle$ is given by 
\begin{eqnarray}
    S = - \Tr \left[ \rho_A \log_2 \rho_A \right] = - \sum_i \lambda_i \log_2 \lambda_i, 
\end{eqnarray}
where we have divided the system into two parts, $A$ and $B$, where $\rho_A = \Tr_B\left(\rho_{AB}\right)$ is the reduced density matrix of subsystem $A$, and $\rho_{AB} = |\psi\rangle\langle\psi|$ is the density matrix of the full state. In the Hermitian case, the distribution of entanglement with energy often unveils interesting physics related to the model. However, in non-Hermitian systems, as the right and left eigenstates can be structurally very different from each other, there are multiple choices for defining the density matrix. They are given by

\begin{eqnarray}
    &(i)&(\rho_k)^{RR} = |R_k\rangle\langle R_k|, \nonumber \\
    &(ii)& (\rho_k)^{LL} = |L_k\rangle\langle L_k|, \nonumber \\
    &(iii)& (\rho_k)^{RL} = |R_k\rangle\langle L_k|.
\end{eqnarray}
The first example corresponds to the density matrix constructed from the $k$th right eigenstate $|R_k\rangle$, while the second corresponds to that constructed from the $k$th left eigenstate $|L_k\rangle$. The third corresponds to the biorthogonal density matrix constructed from the right and left eigenstates. However, the biorthogonal density matrix can in general have complex eigenvalues, making the usual definition of entanglement entropy less straightforward. One commonly considers the absolute values of its eigenvalues inside the logarithm. Since our main motivation is to compare the entanglement properties of the right and left eigenstate ensembles, we do not use the biorthogonal density matrix. Instead, we use $(i)$ and $(ii)$ to construct the density matrices and calculate the entanglement entropy separately for the right and left ensembles.\\

\begin{figure}[htbp]
    \centering
    \includegraphics[width=0.4\textwidth]{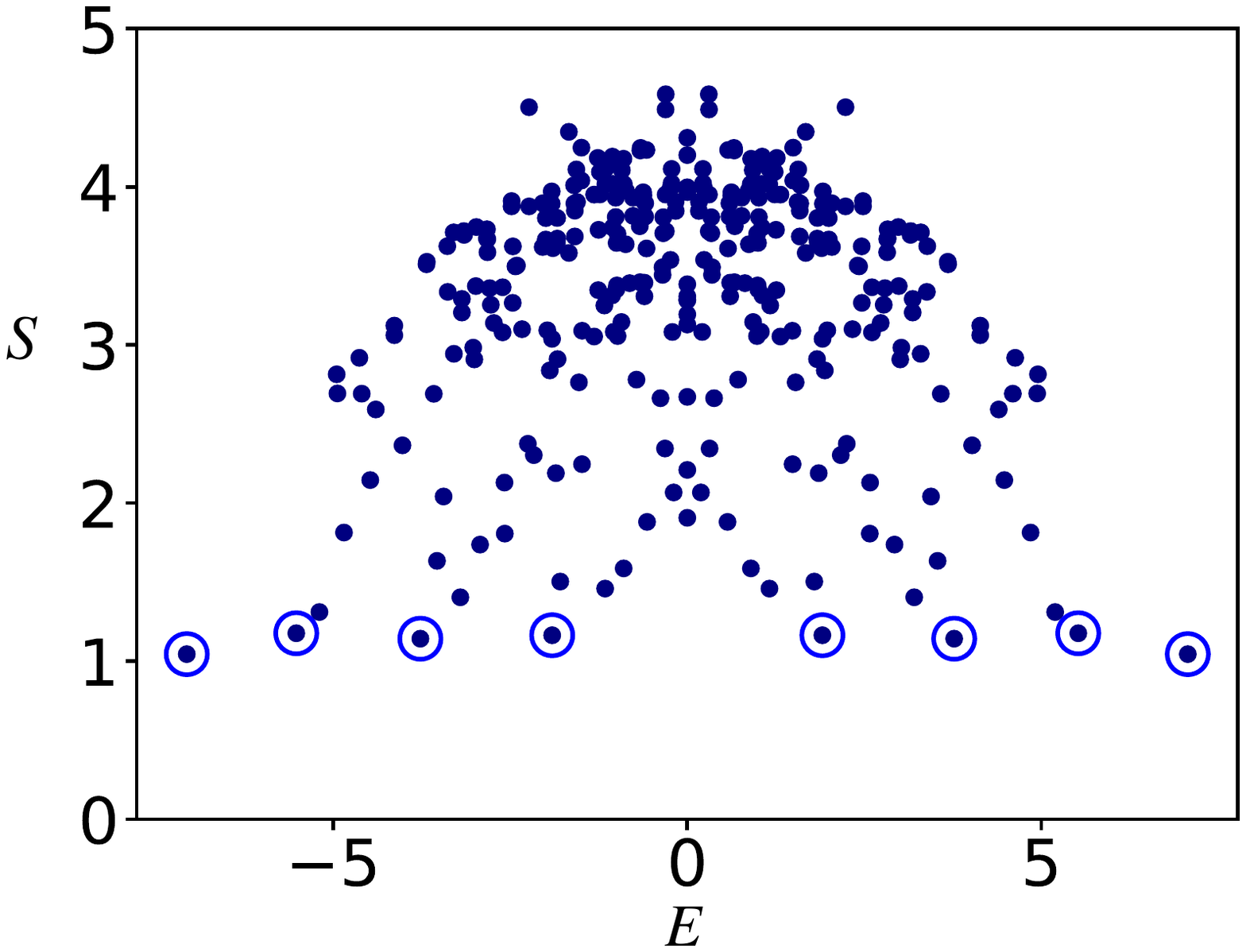}
\caption{\textbf{Entanglement entropy of the left eigenstates in the $\pi$ momentum sector in the non-Hermitian PXP model.} Here we have plotted the entanglement entropy of all the left eigenstates belonging to the $\pi$ momentum sector at non-Hermiticity strength $\gamma=0.7$. We have marked the scar states by blue circles. We can see that the scar states here are the lowest entanglement entropy states. Here, the system size is $L=18$.}
    \label{fig:pi momentum sector scars}
\end{figure}

\begin{figure*}[htbp]
    \centering
    \includegraphics[width=0.75\textwidth]{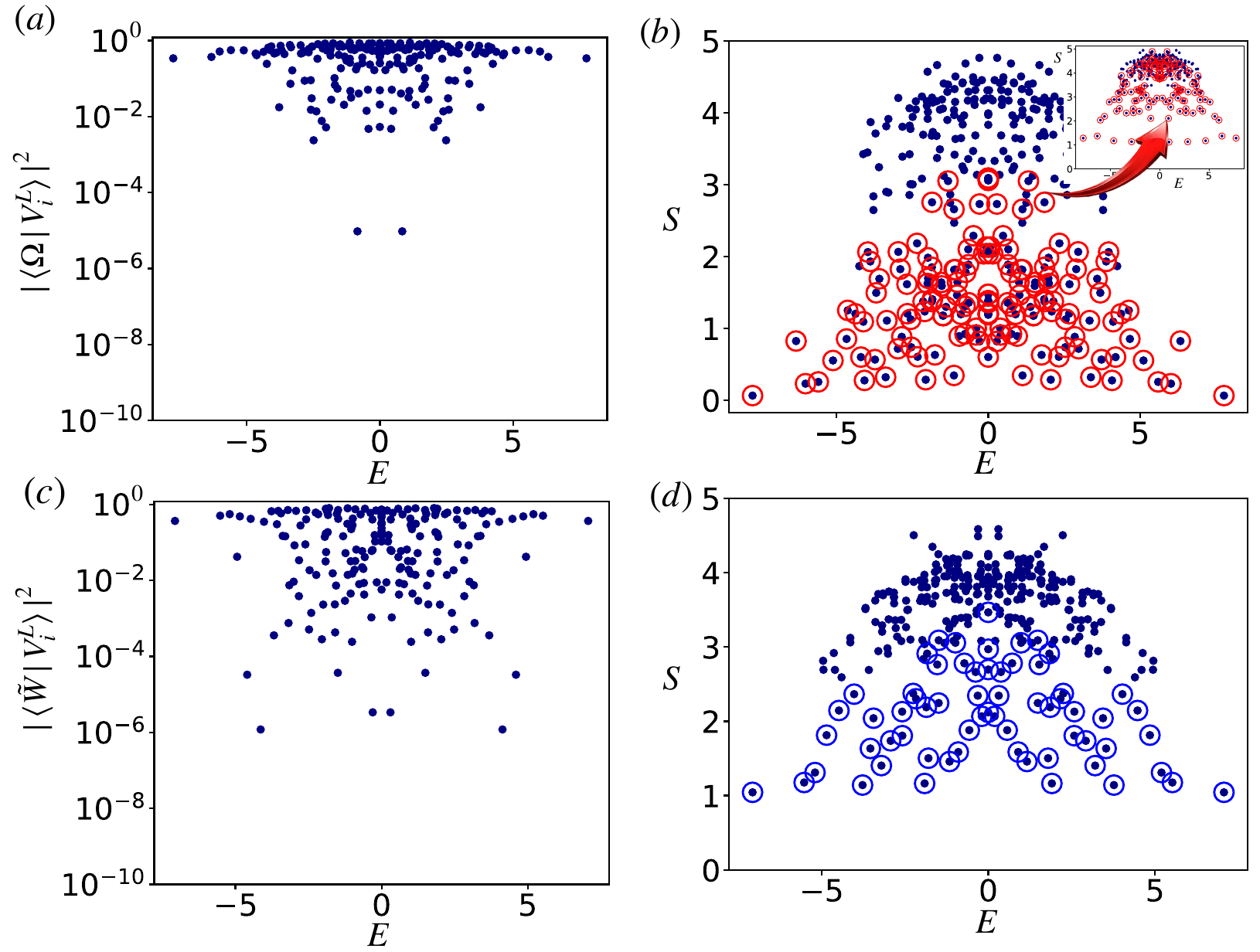}
\caption{\textbf{Entanglement and overlap of left eigenstates in different momentum sectors of the non-Hermitian PXP model.}
(a) Overlap of all left eigenstates in the zero-momentum sector with the state
$|\Omega\rangle=|000\cdots0\rangle$. A substantial number of states exhibit a large overlap with $|\Omega\rangle$.
(b) Entanglement entropy (EE) versus energy $E$ for the same states shown in (a). The states with overlap $\geq 0.33$ with $|\Omega\rangle$ in (a) are marked by red circles in (b), showing a clear correspondence between large overlap with $|\Omega\rangle$ and low entanglement entropy. Additionally, the inset shows the EE of the corresponding states in the right ensemble, using the same eigenstate indices as those identified by red circles in the main plot, (b). In contrast to the left ensemble, these states generally exhibit significantly higher entanglement entropy.
(c) Overlap of all left eigenstates in the $\pi$-momentum sector with the state
$|\tilde{W}\rangle_L=\frac{1}{\sqrt{L}}(|100\cdots0\rangle-|010\cdots0\rangle+\cdots-|000\cdots1\rangle)$.
(d) Entanglement entropy versus energy for the same states shown in (c). The states with overlap $\geq 0.33$ with $|\tilde{W}\rangle_L$ in (c) are marked by blue circles in (d). All plots are for $\gamma=0.7$ and $L=18$.}
    \label{fig:zero state overlap}
\end{figure*}

 In Fig.~\ref{fig:NHPXP_scar}(a)-(c), we show the entanglement entropy plot for all the eigenstates in the Hermitian case and the non-Hermitian right and left ensembles, respectively. In Figs.~\ref{fig:NHPXP_scar}(d)-(f), we show the overlap of all eigenstates with the $|Z_2\rangle$ state for the same systems. We have taken the $k=0$ and $k=\pi$ momentum sectors only. As explained before, all eigenstates can be divided into different momentum sectors fixed by momentum $k=\frac{2\pi n}{L}$. However, for the $|Z_2\rangle$ state, as we have a period of $2$ lattice sites, we get $\mathcal{T}^{2} |Z_2\rangle = |Z_2\rangle$. This gives $e^{i2k} = e^{i 2 \pi m}$ ($m$ integer), yielding  $k= m\pi$. This is why, among all possible momentum sectors, scars are included in the $k=0$ and $\pi$ momentum sectors. Scars belonging to the $k=0$ sector are marked with red circles, and those in the $k=\pi$ sector are indicated by blue circles. In the PXP model with system size $L$, there are a total of $L+1$ scar eigenstates, among which $L$ are nonzero-energy scar states. Moreover,  when $L/2$ is even, there are $L/2$ scars in the zero-momentum sector and $L/2$ scars  in the $\pi$-momentum sector. Similarly,  when $L/2$ is odd, there are  $L/2 +1$ scars in the zero-momentum sector and $L/2 -1$ scars in the $\pi$-momentum sector. For example, in our case, we have taken $L=18$ and there are $10$ scars in the zero-momentum sector and $8$ in the $\pi$-momentum sector.

 In the case of the non-Hermitian right ensemble, the entanglement entropy of
the scar states, shown in Fig.~\ref{fig:NHPXP_scar}(b), is more strongly
suppressed compared to the corresponding scar states in the Hermitian PXP
model. At the same time, these states exhibit a larger overlap with the
$|Z_2\rangle$ state, as shown in Fig.~\ref{fig:NHPXP_scar}(e). This indicates that the scar states in the non-Hermitian right ensemble are more strongly localized around the $|Z_2\rangle$ state and exhibit lower
entanglement compared to their Hermitian counterparts. This is consistent with the
enhanced revival behavior observed in the Krylov dynamics. A qualitatively different behavior is observed for the left ensemble. In this case, the overlap of the scar states with $|Z_2\rangle$ becomes very small,
as shown in Fig.~\ref{fig:NHPXP_scar}(f), indicating that these scar states
are no longer well captured by the $|Z_2\rangle$ state. Interestingly, despite
this significantly reduced overlap, the entanglement entropy of these states
remains low, as shown in Fig.~\ref{fig:NHPXP_scar}(c). Thus, low entanglement
alone continues to identify these states as atypical eigenstates in the left
ensemble, even though their overlap with the conventional scar-inducing
$|Z_2\rangle$ state is strongly suppressed. This suggests that, in the
non-Hermitian setting, low entanglement and large overlap with $|Z_2\rangle$
need not remain correlated across the different ensembles. Interestingly, in the left ensemble, we also find several additional
low-entanglement eigenstates apart from the identified scar states, which are
not present in either the Hermitian or the right ensemble. We believe that these characteristics warrant a more detailed discussion, which we provide below.

\begin{figure*}
    \centering
    \includegraphics[width=  \textwidth]{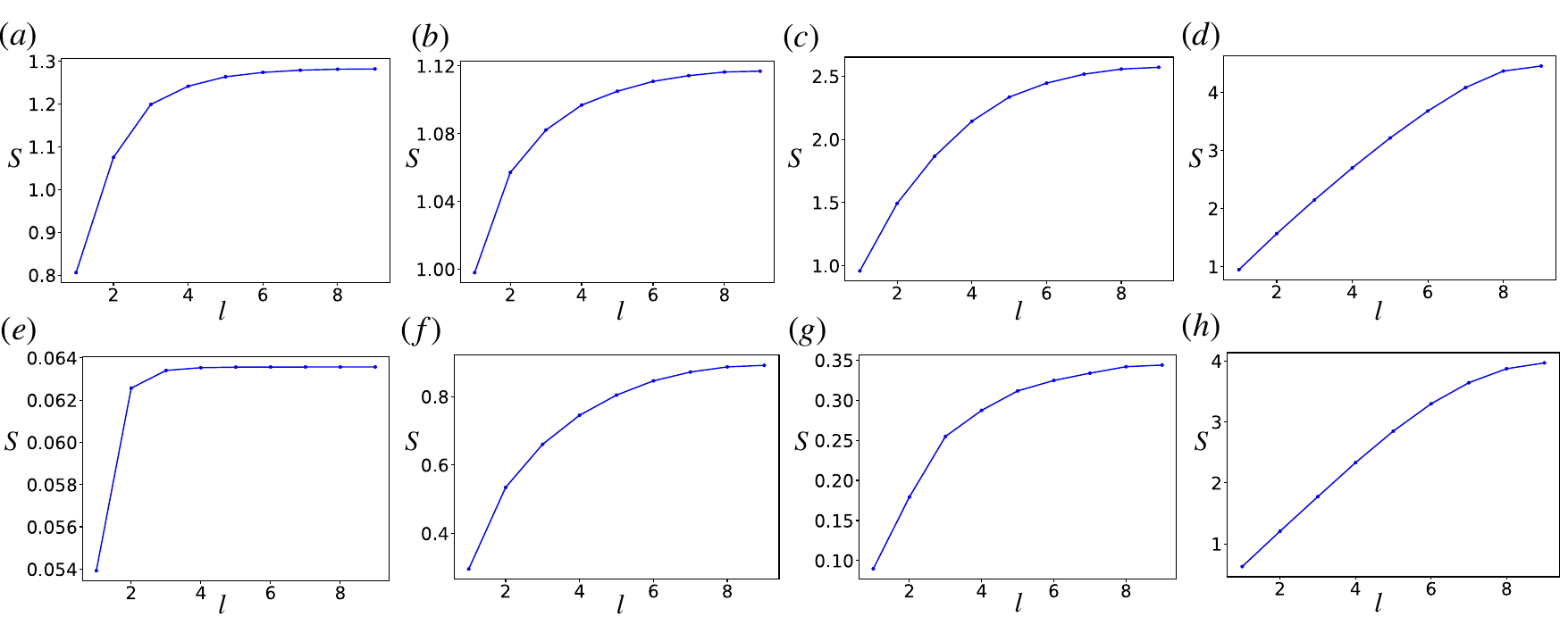}
\caption{{\bf Entanglement entropy $(S)$ scaling with subsystem size $l$.} All plots are obtained for $L=18$ and $\gamma=0.7$, with the subsystem size ranging from $l=1$ to $9$. (a)--(d) correspond to the right ensemble, while (e)--(h) correspond to the left ensemble. (a) and (e) show the ground state, (b) and (f) show a bulk scar state, and (c) and (g) show a non-scar bulk state with very low entanglement entropy in the left ensemble, along with the corresponding bulk state with the same eigenstate index in the right ensemble. Additionally, (d) and (h) show a bulk random state. The ground states exhibit area-law scaling, while the random bulk states show volume-law scaling. Both the scar states and the non-scar low-entanglement states exhibit sub-volume-law scaling, which is approximately logarithmic.}
    \label{fig:EE_scaling}
\end{figure*}

\subsubsection{Low Entanglement Entropy states in Left Ensemble}

We now discuss the entanglement structure of the left eigenstates. As
shown in Fig.~\ref{fig:NHPXP_scar}(c), the scar states in the
zero-momentum sector exhibit a stronger suppression of entanglement
entropy than those in the $\pi$-momentum sector, giving rise to the
zigzag-like structure of the scar states. More importantly, Fig.~\ref{fig:NHPXP_scar}(c)
reveals a number of additional eigenstates with very low entanglement
entropy that do not belong to the identified scar states. In fact,
some of these states have an entanglement entropy even lower than that
of the scar states. These additional low-entanglement states are
responsible for the deviation from the usual arch-like structure
expected for a chaotic spectrum.

To identify the origin of these states, it is useful to examine the two
momentum sectors separately. In the $\pi$-momentum sector, shown
separately in Fig.~\ref{fig:pi momentum sector scars}, the scar states
form the lowest-entanglement states in the spectrum, and we do not
observe a comparable set of additional low-entanglement states. Thus,
the extra low-entanglement states visible in the full spectrum in
Fig.~\ref{fig:NHPXP_scar}(c) originate entirely from the
zero-momentum sector. This motivates us to first examine the
zero-momentum sector in more detail.

In the strong non-Hermiticity limit $\gamma\rightarrow1$, the vacuum
state
\[
|\Omega\rangle=|000\cdots0\rangle
\]
becomes an exact zero-energy left eigenstate of the non-Hermitian PXP
model. This suggests that the additional low-entanglement states in the
zero-momentum sector may be associated with their overlap with
$|\Omega\rangle$. To investigate this, in Fig.~\ref{fig:zero state overlap}(a)
we plot the overlap of all left eigenstates in the zero-momentum sector
with $|\Omega\rangle$, while Fig.~\ref{fig:zero state overlap}(b)
shows the entanglement entropy of the same eigenstates. The states
with overlap $\geq 0.33$ with $|\Omega\rangle$ in Fig.~\ref{fig:zero
state overlap}(a) are marked by red circles in Fig.~\ref{fig:zero state
overlap}(b). A clear correspondence is observed: the states with a
large overlap with the product state $|\Omega\rangle$ have strongly
suppressed entanglement entropy. Thus, the additional low-entanglement
states observed in the zero-momentum sector of Fig.~\ref{fig:NHPXP_scar}(c)
are closely associated with a large overlap with $|\Omega\rangle$.

\begin{table*}[t]
\centering
\caption{\textbf{Summary of the entanglement and overlap properties of the scar and atypical states} in the non-Hermitian PXP model.}
\begin{tabular}{|c|c|c|c|c|c|c|}
\hline
\textbf{State} &
\textbf{Ensemble} &
\textbf{Momentum} &
\textbf{EE} &
$|Z_2\rangle$ &
$|\Omega\rangle$ &
$|\tilde W\rangle_L$
\\
\hline
Scar &
Right &
$0$ &
Low &
Large &
Small &
Zero
\\
\hline
Scar &
Right &
$\pi$ &
Low &
Large &
Zero &
Small
\\
\hline
Scar &
Left &
$0$ &
Low &
Small &
Large&
Zero
\\
\hline
Scar &
Left &
$\pi$ &
Low &
Small &
Zero &
Large
\\
\hline
Atypical &
Left &
$0$ &
Very low &
Small &
Large &
Zero
\\
\hline
\end{tabular}

\vspace{2mm}
\label{tab:scar_summary}
\end{table*}

We next consider the $\pi$-momentum sector. In this sector, the
corresponding simple state is the $L$-site state
\[
|\tilde{W}\rangle_L=
\frac{1}{\sqrt{L}}
\left(
|100\cdots0\rangle-|010\cdots0\rangle+\cdots-|000\cdots1\rangle
\right),
\]
which has the appropriate $\pi$-momentum structure. In
Fig.~\ref{fig:zero state overlap}(c), we plot the overlap of all left
eigenstates in the $\pi$-momentum sector with $|\tilde{W}\rangle_L$,
while Fig.~\ref{fig:zero state overlap}(d) shows the corresponding
entanglement entropy. The states with overlap $\geq 0.33$ are marked by
blue circles. In contrast to the zero-momentum sector, the states with
a substantial overlap with $|\tilde{W}\rangle_L$ are concentrated
among the low-entanglement states associated with the scar structure,
consistent with the absence of a separate population of anomalously
low-entanglement states in the $\pi$ sector.

The observed overlap structure can be understood from the strong
non-Hermiticity limit. As $\gamma\rightarrow1$, both
\[
|Z_2\rangle=|1010\cdots\rangle,
\qquad
|Z_2^{\prime}\rangle=|0101\cdots\rangle
\]
become exact zero-energy right eigenstates of the non-Hermitian PXP
model. The left eigenstates, however, develop distinct structures in
the two momentum sectors: the zero-momentum sector is associated with
the product state
$|\Omega\rangle=|000\cdots0\rangle$, while the $\pi$-momentum sector is
associated with the single-excitation state $|\tilde{W}\rangle_L$.
The increasing overlap with these simple states as $\gamma$ approaches
the strong non-Hermiticity regime therefore provides a natural
explanation for the suppressed entanglement observed in the left
eigenstates. Importantly, the $|\Omega\rangle$ structure extends beyond
the identified scar states to several other zero-momentum eigenstates,
giving rise to the additional atypical low-entanglement states
discussed above.

The additional low-entanglement states associated with a large overlap
with $|\Omega\rangle$ are not conventionally identified as quantum
many-body scar states of the PXP model. Nevertheless, they share an
important characteristic with conventional scar states: a substantial
overlap with a simple state is accompanied by suppressed entanglement.
Furthermore, after resolving the parity symmetry, we find that all
these additional low-entanglement states belong to the
parity-symmetric sector. This differs from the conventional PXP scar
states, which occur in both the symmetric and antisymmetric parity
sectors. We therefore refer to these states as ``atypical'' states
rather than conventional scar states. The inset of Figure~\ref{fig:zero state overlap}(b) further shows that these low-entanglement atypical states identified in the left ensemble do not retain the same low-entanglement character in the corresponding right ensemble, when states with the same eigenstate indices are compared. This highlights special low entangled non-scar bulk states that is characteristics of the left ensemble alone. In Table~\ref{tab:scar_summary}, we summarize the main observations discussed above.

\begin{figure*}
    \centering
    \includegraphics[width=0.8\textwidth]{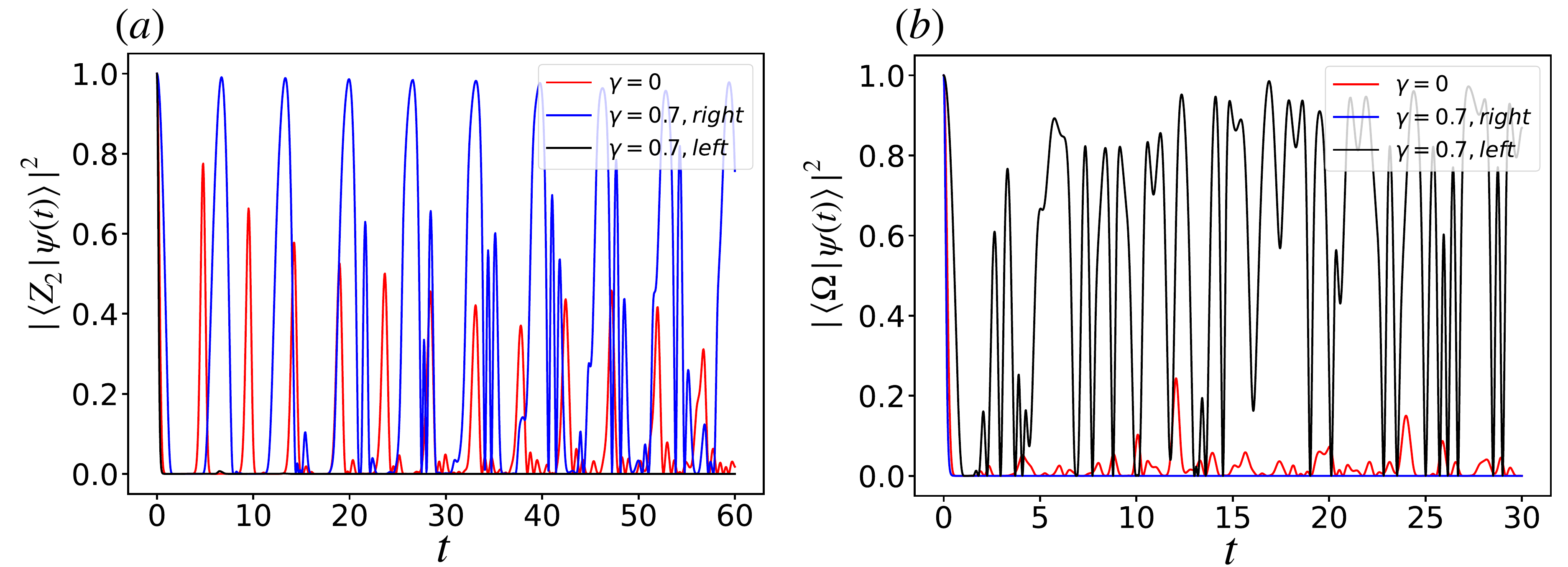}
 \caption{{\bf Return probability of the time-evolved state in the non-Hermitian PXP model.} In (a), we show the fidelity, or return probability, $|\langle\psi(t)|\psi(0)\rangle|^2$ for the initial state $|\psi(0)\rangle=|Z_2\rangle$. The right ensemble (blue) shows almost perfect revivals, whereas the left ensemble shows no appreciable revival. In (b), we take $|\psi(0)\rangle=|\Omega\rangle$ as the initial state. In contrast to (a), the left ensemble (black) exhibits pronounced revivals, while the revival in the right ensemble is strongly suppressed. We use $L=18$ and $\gamma=0.7$ for all non-Hermitian cases.}
    \label{fig:fidelity}
\end{figure*}

In addition to the fixed-system-size entanglement analysis, to better understand the structure of the  scar states in the left ensemble, we examine their entanglement entropy scaling with system size. For both ensembles, we consider four representative states: (i) the ground state, (ii) a bulk scar state, (iii) a bulk non-scar state with low entanglement in the left ensemble and its corresponding state with the same index in the right ensemble, and (iv) finally a bulk state exhibiting high thermal entanglement. The corresponding results are shown in Fig.~\ref{fig:EE_scaling}, with panels (a)--(d) for the right ensemble and (e)--(h) for the left ensemble. The ground states in panels (a) and (e) show area-law behavior, while the scar states exhibit sub-volume entanglement scaling. In contrast, the bulk states with high thermal entanglement, shown in panels (d) and (h), display the expected volume-law behavior. Interestingly, the bulk non-scar low-entanglement states shown in panels (b) and (f) also exhibit sub-volume entanglement scaling, similar to the scar states. Moreover, in the left ensemble, their entanglement entropy is substantially lower than that of the corresponding states in the right ensemble. Thus, the anomalously low entanglement of these non-scar states is not restricted to a fixed system size but persists in their scaling behavior with system size.

\subsubsection{Fidelity with probe states}

The above analysis of the scar states, particularly their entanglement entropy and overlaps with different reference states, motivates us to further examine their dynamical signatures through fidelity. Let $|\psi(0)\rangle$ be an initial state. The time-evolved state at time $t$ is denoted by $|\psi(t)\rangle$. We compute the fidelity
$|\langle\psi(0)|\psi(t)\rangle|^2$ and study its dynamics as a function of $t$. To study the behavior in the right ensemble, we time evolve the initial state using the Hamiltonian $H_{\mathrm{NH}}$, while for the left ensemble, we use $H_{\mathrm{NH}}^{\dagger}$. Since the time evolution generated by a non-Hermitian Hamiltonian is no longer unitary and therefore does not preserve the norm of the state, we normalize the time-evolved states at each time step before computing the fidelity.

In Fig.~\ref{fig:fidelity}(a), we take $|\psi(0)\rangle=|Z_2\rangle$ as the initial state and compute the dynamics of $|\langle\psi(0)|\psi(t)\rangle|^2$. We see that at high non-Hermiticity, the right ensemble (blue) shows an almost perfect revival, whereas in the left ensemble (black), the revival almost completely dies out. We also included the Hermitian case (red) as a reference. This behavior can be understood from the scar structure discussed in the previous section: in the right ensemble at high non-Hermiticity, the scar states have a higher overlap with the $|Z_2\rangle$ state, while this overlap becomes very small in the left ensemble. However, as discussed before, at high $\gamma$, many states in the left ensemble start to show a large overlap with $|\Omega\rangle$ state. We therefore also examine the fidelity starting from this state as depicted in  Fig.~\ref{fig:fidelity}(b). In contrast to Fig.~\ref{fig:fidelity}(a), the left ensemble (black) now shows a large revival, whereas in the right ensemble (blue), the fidelity becomes almost zero at very early times and remains so at later times. Here, we also include the Hermitian case (red) for reference. 

Interestingly, apart from these standard probe states, we also construct a different state starting from the $|Z_2\rangle$ state that shows larger revivals in the Hermitian PXP model than the $|Z_2\rangle$ state itself. This is discussed in Appendix~\ref{sec:hermitian_fidelity}.

\subsection{Entanglement and Magnetization Dynamics}

\begin{figure*}
    \centering
    \includegraphics[width=0.85\textwidth]{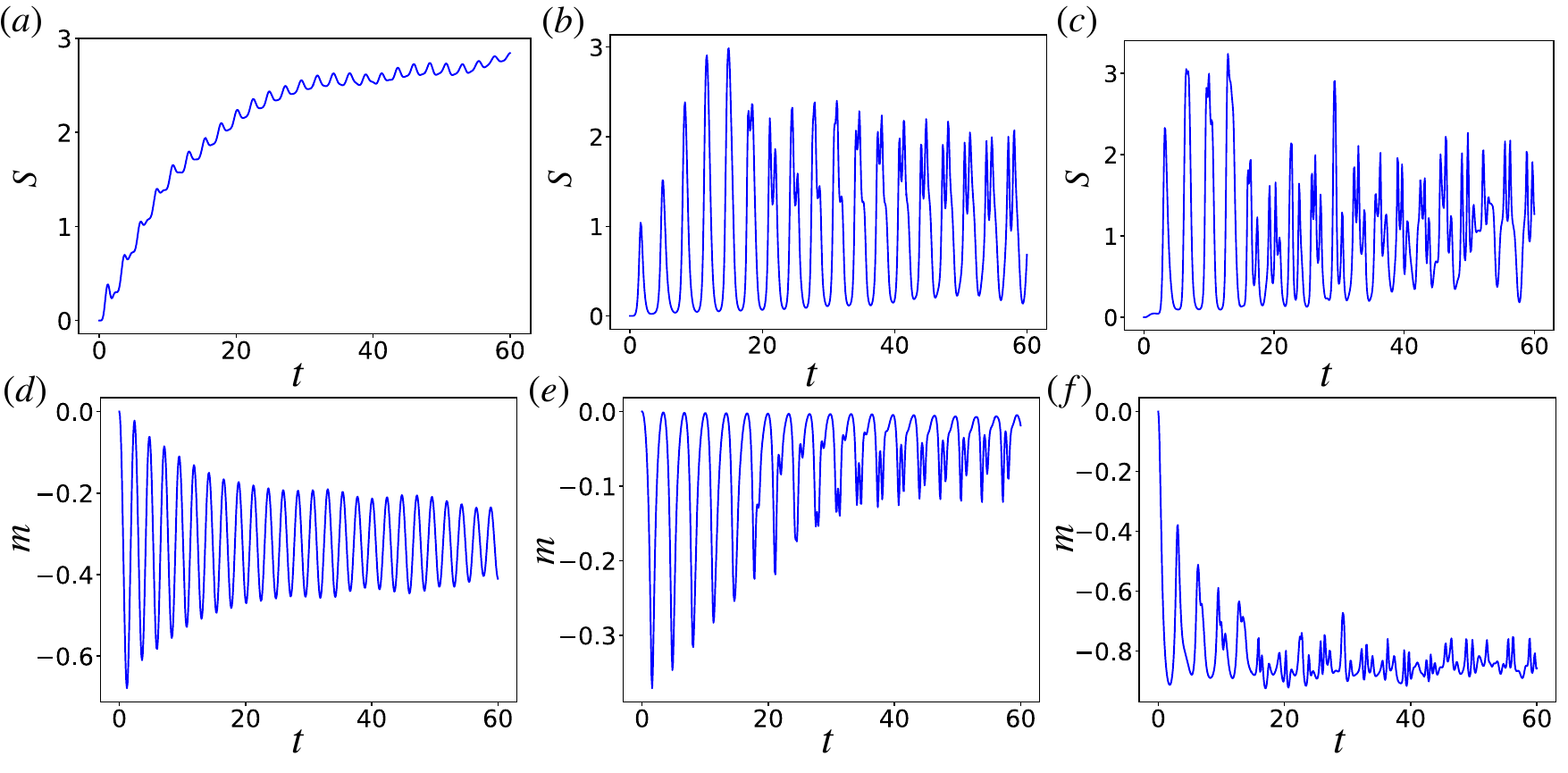}
    \caption{{\bf Entanglement and magnetization dynamics of the left and right ensembles.} (a)-(c): Entanglement dynamics in Hermitian, non-Hermitian right and non-Hermitian left ensembles respectively. (d)-(f): magnetization dynamics in same three ensemble. In case of non-Hermitian, $\gamma$ is $0.7$. All plots are for $L=18$. The distinct behaviour in left and right ensemble is clearly visible from the plots.}
    \label{fig:EE_vs_t}
\end{figure*}

We complete our analysis by examining the dynamics at the level of observables, where the two ensembles show quite distinct behavior. In particular, we study two quantities, the entanglement entropy $S$ and the magnetization (as defined below), for both ensembles and make a comparative study. The magnetization is given by
\begin{equation}
m = \left\langle \frac{1}{L}\sum_{i=1}^L \sigma_i^z \right\rangle.
\end{equation}
 As before, for the right ensemble, we time evolve the initial state 
$|Z_2\rangle$ using $H_{\mathrm{NH}}$, whereas for the left ensemble, we use $H^{\dagger}_{\mathrm{NH}}$. The Hermitian case is included as a reference to understand the effect of non-Hermiticity.
 In Fig.~\ref{fig:EE_vs_t}(a)-(c), we study the dynamics of the entanglement entropy. We compare the entanglement generation starting from the same initial state in three different cases: the Hermitian case (Fig.~\ref{fig:EE_vs_t}(a)), the non-Hermitian right ensemble (Fig.~\ref{fig:EE_vs_t}(b)), and the non-Hermitian left ensemble (Fig.~\ref{fig:EE_vs_t}(c)). We observe that, unlike the steady growth in the Hermitian case, the entanglement in both non-Hermitian ensembles behaves differently and exhibits strong oscillations. Similarly, when we study the dynamics of the magnetization in Fig.~\ref{fig:EE_vs_t}(d)-(f) for the Hermitian, non-Hermitian right, and non-Hermitian left ensembles, respectively, we observe some distinct signatures. In the case of the right ensemble, the magnetization oscillates around  nearly at zero, whereas for the left ensemble, the mean remains close to $-1$. In all non-Hermitian cases, the non-Hermiticity strength is set to $\gamma=0.7$.

The distinct behavior of the two ensembles is particularly interesting in light of the fidelity results discussed above. While the fidelity measures the overall return of the state to its initial configuration, the entanglement and magnetization dynamics provide information about how the state spreads through the many-body Hilbert space and how its physical properties evolve. Thus, these quantities can reveal differences between the left and right ensembles that are not apparent from fidelity alone.

\section{Summary And Discussion}
\label{sec:discussion}

In this article, we have studied non-Hermitian generalizations of the paradigmatic PXP and PPXPP models, with particular emphasis on the differences between their left and right ensembles. For the non-Hermitian PXP model, we find a pronounced left--right asymmetry in both Krylov and eigenstate properties. Using the Arnoldi algorithm, we showed that for the $|Z_2\rangle$ Krylov seed, the right-ensemble Krylov complexity retains oscillatory behavior reminiscent of the Hermitian model, with an increasing oscillation period as the non-Hermiticity $\gamma$ is increased. In contrast, the left-ensemble complexity approaches an almost constant value with fluctuations. The corresponding Arnoldi coefficients further distinguish the two ensembles: the characteristic arch structure is progressively destroyed in the right ensemble, while it survives in the left ensemble, although its width decreases with $\gamma$. These observations raise the question of whether an underlying $SU(2)$ structure, which is known to play an important role in the Hermitian PXP model, can be identified or generalized in the non-Hermitian setting.

We also find a qualitatively different organization of atypical eigenstates in the two ensembles. In the right ensemble, the conventional scar states remain characterized by low entanglement entropy and enhanced overlap with the $|Z_2\rangle$ state. In the left ensemble, the scar states continue to have low entanglement, but their overlap with $|Z_2\rangle$ is strongly suppressed. More importantly, several additional zero-momentum, inversion-symmetric eigenstates exhibit exceptionally low entanglement entropy and large overlap with the product state $|\Omega\rangle=|000\ldots0\rangle$. We refer to these states as atypical states rather than conventional scars. Thus, low entanglement survives as a signature of atypical eigenstates in the left ensemble, while its connection to the conventional scar reference state is lost.

This different overlap structure is also reflected directly in the dynamics. An initial $|Z_2\rangle$ state produces pronounced revivals in the right ensemble, whereas an initial $|\Omega\rangle$ state produces pronounced revivals in the left ensemble. The choice of the initial state therefore strongly affects which part of the non-Hermitian spectrum is dynamically accessed. Our results demonstrate that left and right eigenstate ensembles can encode qualitatively different notions of complexity, scarring, and dynamical revival in the same non-Hermitian Hamiltonian.

Finally, we find corresponding differences in magnetization and entanglement dynamics and extend our analysis to the non-Hermitian PPXPP model. As an open problem, it would be interesting to explore whether these distinct dynamical responses can be useful for quantum technology applications in particular quantum sensing and parameter estimation. 

\acknowledgments{S.S.R. acknowledges financial support from the Faculty Research Scheme, IIT (ISM) Dhanbad, India, under Project No. FRS/2024/PHYSICS/MISC0110, and from the Anusandhan National Research Foundation (ANRF), Government of India, under Grant Nos. ANRF/ARG/2025/004617/PS and ANRF/ECRG/2025/002793/PMS. We also acknowledge the use of the QuSpin package~\cite{QuSpin} for the numerical simulations presented  in this work. }


\appendix

\begin{widetext}

\section{Results for PPXPP Model}
\label{PPXPP}

\begin{figure*}
\centering
\includegraphics[width=0.9\textwidth]{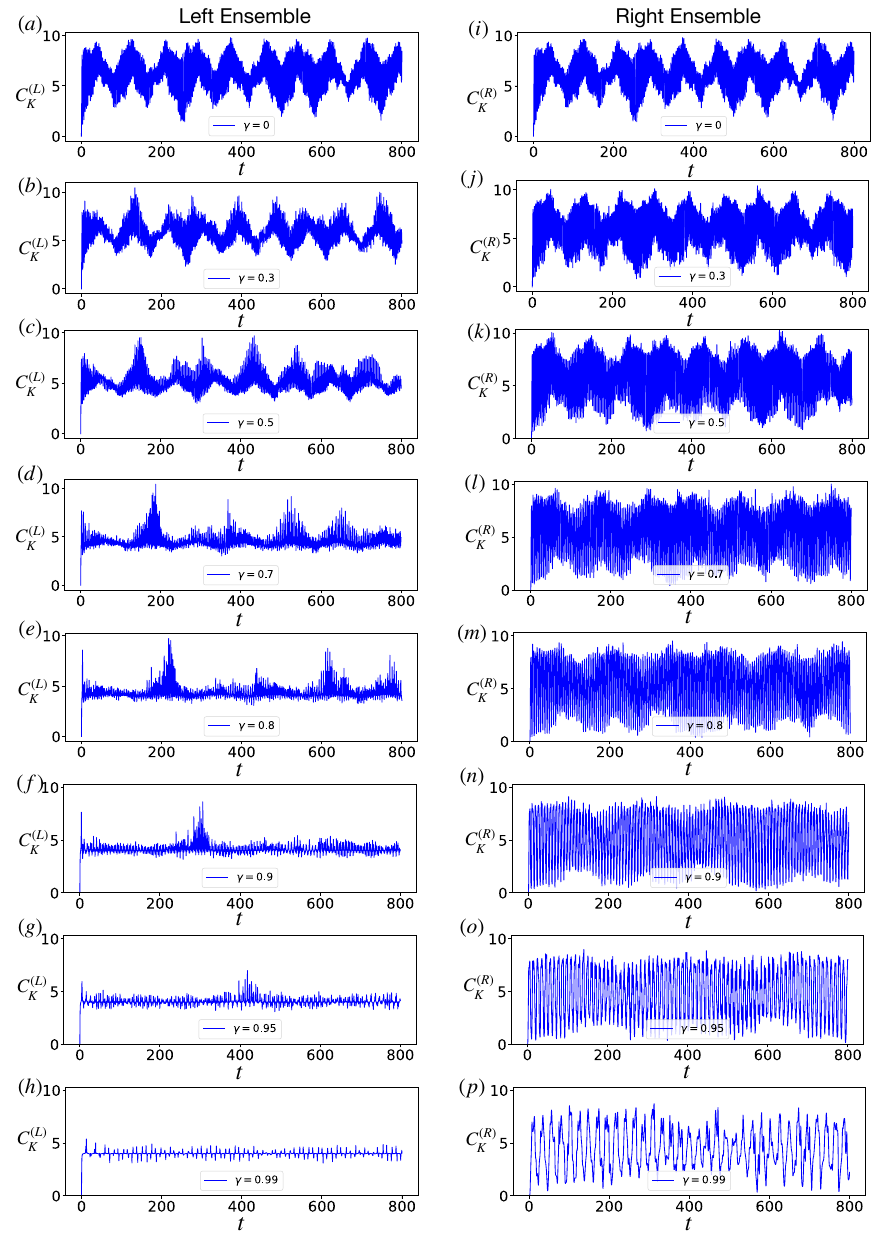}
\caption{\textbf{Krylov complexity in the non-Hermitian PPXPP model.} (a)--(h) correspond to the left ensemble, while (i)--(p) correspond to the right ensemble. The plots show the time evolution of the Krylov complexity for different values of the non-Hermiticity parameter $\gamma$. In the left ensemble, the Krylov complexity gradually saturates with increasing $\gamma$, whereas in the right ensemble, it develops increasingly pronounced oscillations. Thus, the two ensembles exhibit qualitatively distinct behaviour under non-Hermitian evolution. All plots are for $L=16$ and the initial state is $|K=4\rangle$.}
\label{fig:NHPPXPP_K_complexity}
\end{figure*}

The PPXPP model, an extension of the PXP model, is also known to exhibit quantum many-body scars. However, the $|Z_2\rangle$ state which is used in PXP model, is not a good choice for investigating the scar structure in here. Rather, a different state is used here. The state used to probe the scar structure of this model is given by~\cite{PPXPP}
\begin{eqnarray}
|K\rangle = \bigotimes_{i=1}^M \left( |0\rangle_{K/2} \otimes |W\rangle_{K/2} \right).
\end{eqnarray}
Here, $|0\rangle_{K/2}$ denotes $|00...0\rangle$ with $K/2$ zeros, and $|W\rangle_{K/2}$ is given by
\begin{eqnarray}
|W\rangle_{K/2} = \sqrt{\frac{2}{K}} \left( |10...0\rangle + |01...0\rangle + \cdots + |00...1\rangle \right),
\end{eqnarray}
where each state is of length $K/2$. $M$ and $K$ are chosen such that $M\times K=L$. Thus, we can think of the system as consisting of $M$ repeating unit cells, each of length $K$, making up the total lattice of size $L$. The number of energy values at which the scars occur is known to be $2L/K+1$. Since the state $|K\rangle$ has a periodicity of $K$, similar to the analysis of the PXP model with $|Z_2\rangle$, we can identify the definite momentum sectors to which the quantum many-body scars belong. We must have
\begin{eqnarray}
\mathcal{T}^K |K\rangle = |K\rangle,
\end{eqnarray}
 where $\mathcal{T}$ is the translation operator for one lattice site. This gives $e^{ikK}=e^{i2\pi m}$, and hence
\begin{eqnarray}
k=\frac{2\pi m}{K}.
\end{eqnarray}
It turns out that all scar states have $k$ values given by this equation. Specifically, there are $L/K+1$ scars in the $k=0$ sector, $L/K$ scars in the $k=2\pi/K$ sector, and $L/K$ scars in the $k=-2\pi/K$ sector. Thus, there are a total of $3L/K+1$ scar states distributed among $2L/K+1$ energy values. The scars in the $k=\pm 2\pi/K$ sectors occur as degenerate pairs.

Conventional studies~\cite{PPXPP} have used the $|K=6\rangle$ state as a probe. However, here we use the $|K=4\rangle$ state as our probe state. The $|K=4\rangle$ state helps to identify a larger number of scar states than the $|K=6\rangle$ state. In all our studies related to the PPXPP model, we use $L=16$ unless otherwise specified. For the probe state, we take $|K=4\rangle$, which automatically fixes $M=4$.

\subsection{Krylov complexity}
Here, we first discuss the Krylov complexity in the non-Hermitian PPXPP model, shown in Fig.~\ref{fig:NHPPXPP_K_complexity}. In Fig.~\ref{fig:NHPPXPP_bn}, we show the behaviour of $b_n$ versus $n$ for both ensembles. All plots are for $L=16$, and we have taken the $|K=4\rangle$ state as our probe or initial state. Similar to the PXP model, here also we see that, for the right ensemble, the Krylov complexity shows oscillations with increasing $\gamma$, while for the left ensemble, it gradually saturates. In Fig.~\ref{fig:NHPPXPP_bn}, we see that for $\gamma=0$, the arch extends up to $n=8$, which is the number of nonzero energy unique non-degenerate scar states here, since $\frac{2L}{K} = \frac{2\times16}{4}=8$. For the right ensemble, the arch shape vanishes with increasing non-Hermiticity, whereas for the left ensemble, the size of the arch reduces and becomes fixed at $n=4$ for sufficiently large $\gamma$.

\begin{figure}
\centering
\includegraphics[width=0.65\textwidth]{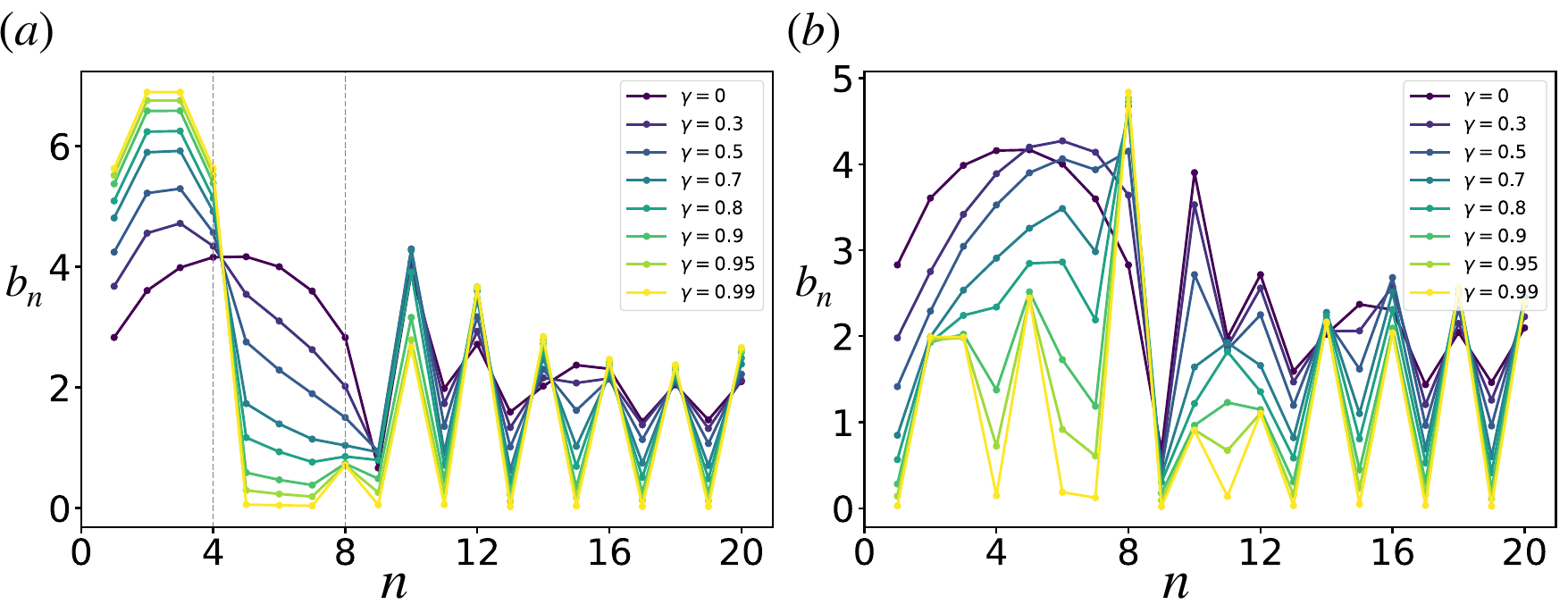}
\caption{\textbf{$b_n$ vs $n$ for the right and left ensembles with varying $\gamma$.} In (a), we show the left ensemble, while in (b), we show the right ensemble. For $\gamma=0$, both ensembles show an arch-like structure that extends up to $n=8$, corresponding to the number of non-zero-energy scar states in the model. With increasing $\gamma$, the arch structure is gradually suppressed in the right ensemble and eventually disappears. In contrast, for the left ensemble, the extent of the arch gradually decreases and, at sufficiently large $\gamma$, becomes fixed at $n=4=8/2$. Thus, the two ensembles show qualitatively different evolution of the Krylov coefficients with increasing non-Hermiticity, similar to the behaviour observed in the non-Hermitian PXP model.}
\label{fig:NHPPXPP_bn}
\end{figure}

\subsection{Scars}
In Fig.~\ref{fig:NHPPXPP_scars}, we show the quantum many-body scars in the non-Hermitian PPXPP model. Here, we have taken $L=20$ and considered only the $k=0$ and $k=\pi/2$ momentum sectors. The number of unique non-zero-energy scar states here is 10. They alternate between the two momentum sectors. We have marked the scars in the zero-momentum sector by red circles and those in the $\pi/2$ momentum sector by blue circles. Similar to the non-Hermitian  PXP case, here also we see that, in the non-Hermitian left ensemble (Fig.~\ref{fig:NHPPXPP_scars}(c)), the scar states have more suppressed entanglement entropy compared with the right ensemble (Fig.~\ref{fig:NHPPXPP_scars}(b)). Further, for the left ensemble, the scars in the zero-momentum sector show much lower entanglement entropy than the scars in the $\pi/2$ momentum sector. Figures~\ref{fig:NHPPXPP_scars}(d)-(f) show the overlap of all the eigenstates with the $|K=4\rangle$ state for the Hermitian, non-Hermitian right, and non-Hermitian left ensembles, respectively.

\begin{figure}
\centering
\includegraphics[width=0.85\textwidth]{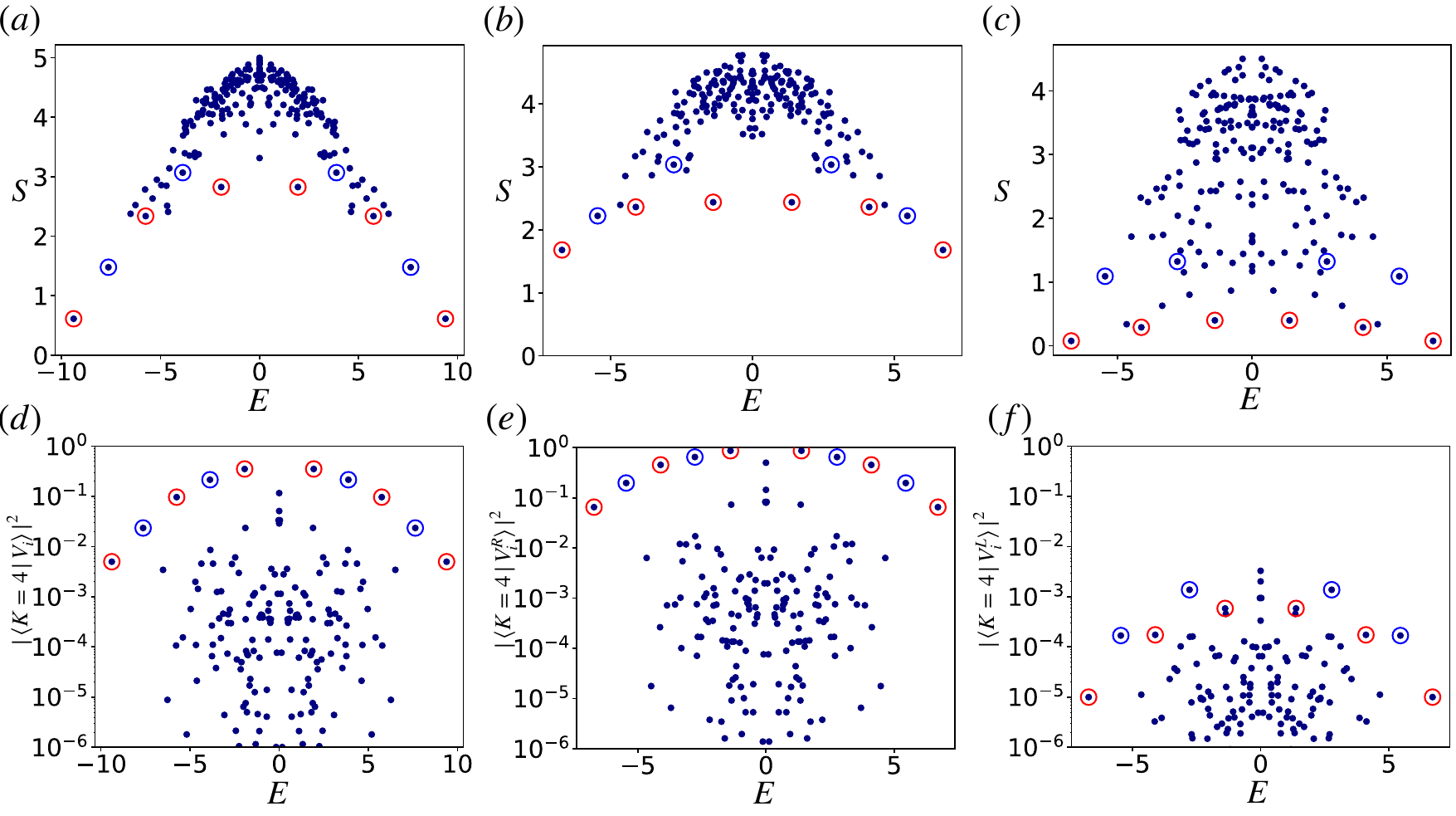}
\caption{\textbf{Entanglement entropy and overlap of eigenstates in the PPXPP model.} (a) EE vs $E$ for the Hermitian $(\gamma=0)$ PPXPP model. (b) Same for the right ensemble of the NHPPXPP model $(\gamma=0.7)$. (c) Same for the left ensemble $(\gamma=0.7)$. (d) Overlap of all eigenstates of the PPXPP model with the $|K=4\rangle$ state. (e) Same for the NHPPXPP model in the right ensemble. (f) Same for the left ensemble. All plots are for $L=20$ and the two momentum sectors $(k=0,\pi/2)$. The scar states are highlighted by circles: red for $k=0$ and blue for $k=\pi/2$. In the left ensemble, the scar states in the two momentum sectors form zig-zag-like structures in the EE vs $E$ plot, with lower entanglement entropy than the corresponding scar states in the right ensemble.}
\label{fig:NHPPXPP_scars}
\end{figure}

\begin{figure}
\centering
\includegraphics[width=0.45\textwidth]{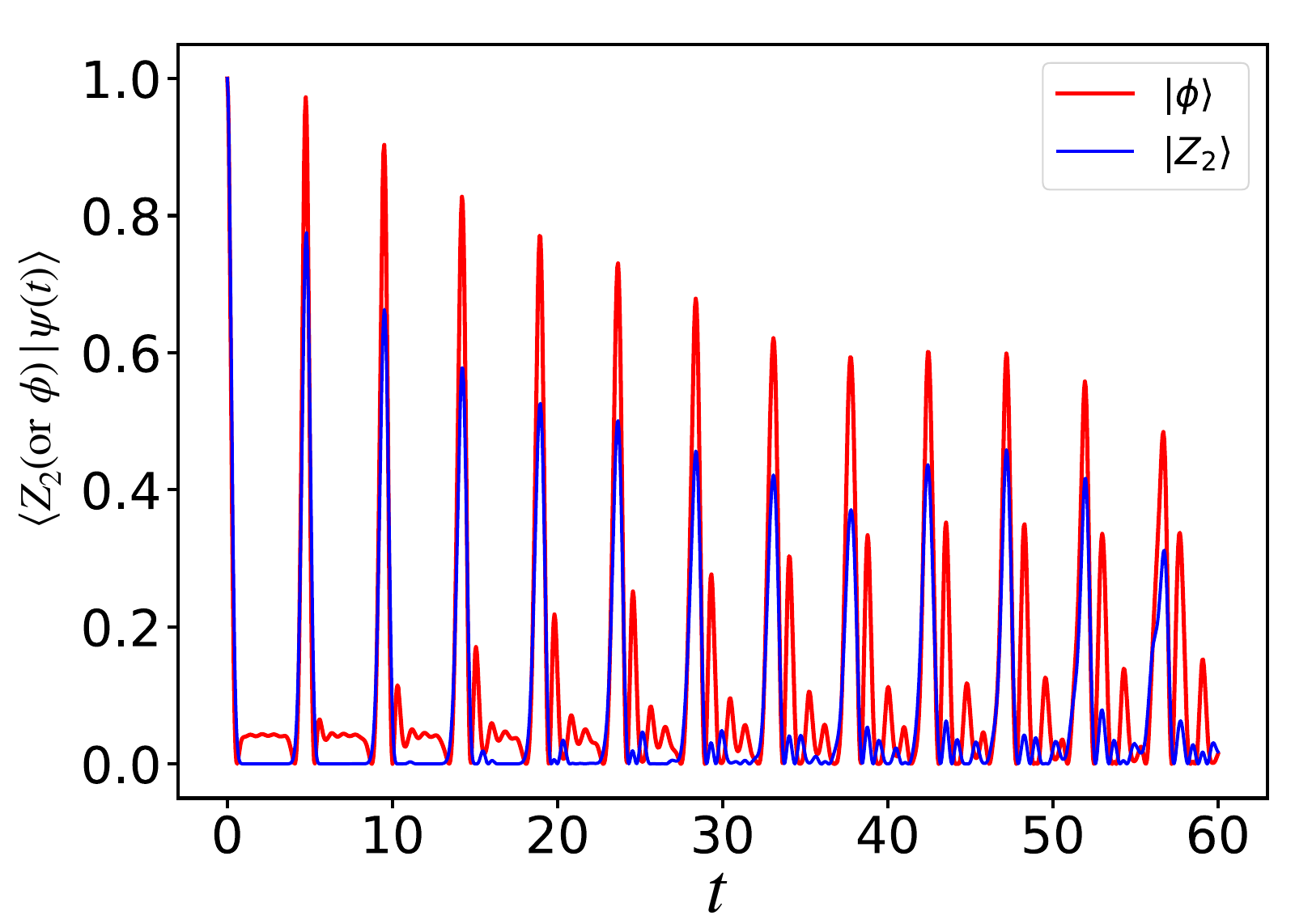}
\caption{{\bf Fidelity in the Hermitian PXP model for various probe states}. The blue line corresponds to the initial state $|Z_2\rangle$, while the red line corresponds to the constructed initial state $|\phi\rangle$.}
\label{fig:Herm_fidelity}
\end{figure}
\section{Enhancing Fidelity in Hermitian PXP Model}
\label{sec:hermitian_fidelity}

In this section, we construct a state $|\phi\rangle$ from the $|Z_2\rangle$ state and use it as the initial state in the fidelity plot. We show that this state gives better fidelity revivals in the Hermitian PXP model. Since the right and left eigenstates are the same in the Hermitian case, we denote them by ${|E_i\rangle}$. Our state $|\phi\rangle$ is given by

\begin{eqnarray}
|\phi\rangle = \frac{|\phi^{\prime}\rangle}{\sqrt{\langle\phi^{\prime}|\phi^{\prime}\rangle}},
\qquad
|\phi^{\prime}\rangle = \sum_n \langle E_n|Z_2\rangle |E_n\rangle .
\end{eqnarray}

\noindent Here, $n$ runs only over the scar eigenstates. Using $|\phi\rangle$ as the initial state, we calculate the fidelity and show the result in Fig.~\ref{fig:Herm_fidelity}. The blue line shows the fidelity starting from $|Z_2\rangle$, while the red line shows the fidelity starting from $|\phi\rangle$. We can clearly see that the constructed state exhibits much larger fidelity revivals than the conventional $|Z_2\rangle$ state.

\end{widetext}

\bibliography{references}{}

\end{document}